\documentclass[a4paper,12pt]{article}
 
\usepackage{bbm}
\usepackage[utf8]{inputenc}
\usepackage{placeins}
\usepackage{amssymb,amsmath,amsfonts}
\usepackage[colorlinks,linkcolor=purple,citecolor=teal]{hyperref}
\usepackage{dsfont}
\usepackage{color}
\usepackage{graphicx}
\usepackage{hyphenat}
\usepackage{wrapfig}
\usepackage{empheq}
\usepackage{textcomp}
\usepackage[caption=false]{subfig}
\usepackage{rotating}
\usepackage{wrapfig}
\usepackage{pdfpages}
\usepackage{verbatim}
\usepackage{setspace}
\usepackage{array}
\usepackage{caption}
\usepackage[pdf]{pstricks}
\usepackage{upgreek}
\usepackage{cmll}
\usepackage{latexsym}
\usepackage{braket}
\usepackage{cite}
\usepackage{epsfig}
\usepackage[left=2.4cm,top=3.3cm,right=2.4cm,bottom=3.3cm,bindingoffset=0cm]{geometry}
\usepackage[titletoc,page]{appendix}
\usepackage{multicol}
\usepackage{youngtab}

\usepackage[normalem]{ulem}     

\newcommand{\beq}{\begin{eqnarray}}
\newcommand{\eeq}{\end{eqnarray}}
\newcommand{\bea}{\begin{eqnarray}}
\newcommand{\eea}{\end{eqnarray}}
\newcommand{\be}{\begin{equation}}
\newcommand{\ee}{\end{equation}}
\newcommand{\diff}{\mathrm{d}}
\newcommand{\tr}{\mathrm{tr}}
\newcommand{\im}{\mathrm{i}}
\newcommand{\calR}{\mathcal{R}}
\newcommand{\rmc}{\mathrm{c}}
\newcommand{\rme}{\mathrm{e}}
\newcommand{\rmf}{\mathrm{f}}

\newcommand{\rmF}{\mathrm{F}}
\newcommand{\rmL}{\mathrm{L}}

\newcommand{\rmS}{\mathrm{S}}
\def\brc{\langle}
\def\ckt{\rangle}
\def\const{{\rm const}}
\def\de{\partial}

\def\nn{\nonumber}

\def\Tr{\qopname\relax o{Tr}}

\numberwithin{equation}{section}

\def\bxi{{\bar{\xi}} }
\def\su{$ \phantom{{{{{\yng(1)}}}}}\!\!\!\!\!\!\!\!$}
\def\sbuu{$\phantom{{{{\bar{\yng(1,1)}}}}}\!\!\!\!\!\!$}
\def\sbu{$\phantom{{\bar{{{\yng(1)}}}}}\!\!\!\!\!\!\!\!$}
\def\sbbuu{$\phantom{{{\bar{\bar{\yng(1,1)}}}}}\!\!\!\!\!\!\!\!$ }
\def\sbbu{ $\phantom{{{\bar{\bar{\yng(1)}}}}}\!\!\!\!\!\!\!\!$ }

\numberwithin{equation}{section}

\begin{document}

\title{
\vskip 20pt
\bf{   Anomalies and  phases   of \\   strongly-coupled
chiral gauge theories:   \\   
recent developments }
}
\vskip 40pt  
\author{  
Stefano Bolognesi$^{(1,2)}$, 
 Kenichi Konishi$^{(1,2)}$, Andrea Luzio$^{(3,2)}$    \\[13pt]
{\em \footnotesize
$^{(1)}$Department of Physics ''E. Fermi", University of Pisa,}\\[-5pt]
{\em \footnotesize
Largo Pontecorvo, 3, Ed. C, 56127 Pisa, Italy}\\[2pt]
{\em \footnotesize
$^{(2)}$INFN, Sezione di Pisa,    
Largo Pontecorvo, 3, Ed. C, 56127 Pisa, Italy}\\[2pt]
{\em \footnotesize
$^{(3)}$Scuola Normale Superiore,   
Piazza dei Cavalieri, 7,  56127  Pisa, Italy}\\[2pt]
\\[1pt] 
{ \footnotesize  stefano.bolognesi@unipi.it, \ \  kenichi.konishi@unipi.it,  \ \  andrea.luzio@sns.it}  
} 
\date{}

\vskip 6pt
\maketitle

\begin{abstract}

After many years of investigations, our understanding of the dynamics of strongly-coupled chiral gauge theories 
is still quite unsatisfactory today.   Conventional wisdom about strongly-coupled gauge theories, successfully applied
 to QCD, is not always as useful in chiral gauge theories. 
Recently some new ideas and techniques have been developed, which involve concepts of generalized symmetries, 
of gauging a discrete center symmetry, and of generalizing the 't Hooft anomaly matching constraints to  
include certain mixed symmetries. This new development has been applied 
 to chiral gauge theories,  leading to many interesting, sometimes quite unexpected, results.
   For instance, in the context of generalized Bars-Yankielowicz and generalized Georgi-Glashow models,  these new types of anomalies give a rather clear indication
   in favor of the dynamical Higgs phase, against confining, flavor symmetric vacua.    
   
  Another closely related topics is strong anomaly and the effective low-energy action representing it. 
  It turns out that they have significant implications on the phase of chiral gauge theories, giving indications consistent with the findings based on
  the generalized anomalies.  
 
 Some striking analogies and contrasts between the massless QCD  and chiral gauge theories seem to emerge from these discussions.      
The aim of this work is to review these developments.

\end{abstract}

\newpage
\tableofcontents

\newpage

\section{Introduction}

One of the mysteries of the world we live in is the fact that it has a nontrivial chiral property. 
 The macroscopic structures such as biological bodies have often approximately left-right symmetric forms, but not exactly.
  At the molecular level,  $O(10^{-6}   \, {\rm cm})$, the structure of DNA possesses 
a definite chiral spiral form. At the microscopic scales of the fundamental interactions, $O(10^{-14} \, {\rm cm})$,
 the left-handed and right-handed quarks and leptons have distinct couplings to the 
 $SU(3)\times SU(2)_L\times U(1)_Y$ gauge bosons. The parity violation in the 
 ''weak interactions" processes, though it was first considered somewhat weird, later found a natural explanation \cite{FeyGel}:  the fundamental entity of our world is 
a set of Weyl fermions in the $\big(\tfrac{1}{2}, 0\big)$ representation of the Lorentz group.  If such fermions are in a generic 
complex representation of the gauge group, the resulting theory will break parity.  In other words, the origin of parity violation is to be traced to the
type of building blocks our world is made of, rather than to a peculiar property of the Fermi interactions.   
 
Most of grand unification schemes such as  those based on $SU(5), SO(10)$ or $E_7$ groups are also all based on chiral gauge theories.    

The Glashow-Weiberg-Salam (GWS) $SU(2)_L\times U(1)_Y$  theory (as well as its GUT generalizations) is a weakly coupled theory and,  as such, is well understood within the framework of perturbation theory. 
But this also means that the 
theory should be regarded, at best, as a very good low-energy effective theory.  In particular, 
it is unlikely that  the gauge symmetry breaking sector 
described by a potential term for the Higgs scalar, though phenomenologically quite successful, is a self-consistent, fundamental description. 
Nevertheless,  attempts to replace it by new, QCD-like strongly-coupled gauge theories (such as Technicolor, Extended technicolor,  Walking technicolor, etc.)  have not been 
entirely successful so far. 
 
On the other hand,   our understanding of {\it strongly-coupled chiral gauge theories} is today uncomfortably limited   \cite{tHooft}-\cite{Caccia}.    This is in a striking contrast to the case of vector-like gauge theories, for which  we have
an extensive  literature. They include 
some general theorems \cite{Vafa:1983tf}, \cite{Vafa:1984xg},  lattice simulations \cite{Debbio:2015ila}-\cite{Athenodorou:2014eua},  the effective Lagrangians \cite{Witten:1983tx}-\cite{AfDiSe},  the conventional  't Hooft anomaly analysis  \cite{tHooft},  the powerful exact results in ${\cal N}=2$ supersymmetric theories \cite{SW1}-\cite{Tachikawa}, space compactification with semi-classical approximation \cite{unsal}-\cite{Unsal:2007vu}, and so on.

These theoretical tools are unfortunately  unavailable for the analysis of strongly-coupled {\it  chiral} gauge theories, with a few exceptions  such as the large $N$ approximation,  the  't Hooft anomaly matching constraints, and  some general considerations based on the renormalization group.   Taken together,   they yield some useful but not very detailed 
knowledge about the dynamics and  phases of the chiral gauge theories.  Partial list of the papers on these efforts are found in \cite{tHooft}-\cite{Caccia}.
 This kind of situation is certainly  limiting rather severely our capability of  finding the place for  chiral gauge theories 
in the context of a realistic theory of the fundamental interactions beyond the standard model, e.g.,  in the context of  composite models for the quarks and leptons, the composite Higgs boson models,    
 the composite dynamical models for dark matter,   and so on. The need for new ideas for making  true progress in this research field is badly felt.  
 
In our view, a  little hope for a breakthrough comes from the very recent ideas involving the concept of generalized symmetries 
\cite{Seiberg}-\cite{GKSW}.     
Pure  Yang-Mills theories and QCD-like theories have been  analyzed  by using 
new,  stronger,  version   of 't Hooft anomaly matching constraints,   involving $0$-form and $1$-form  symmetries together \cite{GKKS}-\cite{BKL2}. 

 The generalized symmetries are symmetries  which  do not act on local field operators, as the conventional symmetries,  
but only on extended objects,  such as closed lines and surfaces. As the corresponding gauge functions are now 1-form, 2-form  fields (in the standard gauging of global symmetries the gauge transformation parameters  which appear in the exponent are just -  0-form -  functions of the spacetime), these new types of symmetries are sometimes called 
1-form-, 2-form-, etc symmetries. 

A  familiar example of the 1-form symmetry is the  ${\mathbbm Z}_N$  center symmetry  in Euclidean  $SU(N)$ Yang-Mills theory at finite temperature, which acts on the Polyakov loop. The unbroken (or broken) center symmetry  by the vacuum expectation value (VEV)  of the Polyakov loop, is a valid criterion for confinement (or de-confinement) phase.  Similarly, the area law / perimeter law  of the  VEV of the Wilson loops can be
considered as the criterion for confinement / Higgs phase.  Note however that the center symmetry ${\mathbbm Z}_N$ as  used this way is a {\it  global} 1-form symmetry. 
The main input of the new development \cite{GKKS}-\cite{BKL2}  is the idea of {\it gauging}   1-form symmetries.  


In fact,   these generalized symmetries {\it are} symmetries of the systems being  considered, even if they act  in a way different from the conventional symmetry action. We are free  to decide to ''gauge" these new types of symmetries.    
Anomalies  one encounters in doing so are obstructions of gauging a symmetry,  which is by definition a 't Hooft anomaly.   
 And as in the usual requirement of the  ultraviolet (UV) - infrared  (IR)   ''anomaly matching",  a similar conditions  arise 
 in gauging the generalized  (higher-form) symmetries together with some conventional  (''0-form") symmetries, which have come to be known in recent literature as a  ''mixed 't Hooft anomaly". 
 As will be seen below,  these constraints carry significant information on the dynamics of wide classes of  chiral gauge theories as well,  which is our main interest.

Very recently,  the present authors have realized \cite{BKL5} that  the strong anomaly, which plays a prominent role in the solution of the so-called $U(1)_A$ problem in QCD,  can also  be significant in the study of the phases of chiral gauge theories, such as those discussed in this review.  
A key observation is that the well-known low-energy strong-anomaly effective action for QCD, which reproduces the 
effects of the strong anomaly in the low-energy effective action, has  a nontrivial implication on the symmetry breaking pattern itself.  For some reason these ideas have not been applied much to the study of strongly-interacting chiral gauge theories until now.
 
It is found that, quite remarkably, the considerations based on the strong anomaly yield  similar indications on the phase of the chiral gauge theories, 
as  those found by applying the generalized 't Hooft anomalies.   Even though they are arguments fully independent of each other,  the agreement of the results should not 
probably be considered  entirely accidental. 
Indeed they  both originate from the proper treatment of the strong anomaly on various  $U(1)$  symmetries present in the theory.

The rest of the work is organized as follows.  In Sec.~\ref{Method1}   the procedure of the computing anomalies associated with {\it  the gauging of certain 1-form discrete symmetry}  is  discussed, as this constitutes one of  the main theoretical tools of our analysis.    
The  detail of the discussion is further  divided in two parts.  The first part,  Sec.~\ref{review},  concerns models  with 1-form center symmetry  ${\mathbbm Z}_k \subset  {\mathbbm Z}_N \subset SU(N)$    which does not act on the matter fermions. These models have ordinary (0-form) discrete symmetries also, call  ${\mathbbm Z}_{\ell}$,   which are nonanomalous remnants of anomalous $U(1)$ symmetries. 
%

The second class of models contain some matter fermions in the fundamental or antifundamental representation of $SU(N)$. 
Normally one would conclude that 1-form center symmetry  ${\mathbbm Z}_N$  is simply absent in such models.  
However, as explained in Sec.~\ref{review2},  it turns out that  it is still possible to define a color-flavor locked  1-form center symmetry  ${\mathbbm Z}_N$.

 In   Sec.~\ref{Simplecases}   and  Sec.~\ref{BYGG},  applications to various chiral gauge theories of these new 't Hooft anomaly  constraints  are explored.  
In Sec.~\ref{Simplecases},  physics of various chiral gauge theories of the first kind are discussed, by using the general results of Sec.~\ref{review}. 
 Sec.~\ref{BYGG}  is dedicated to  the applications of the formulas  found in   Sec.~\ref{review2}  
  to  two large classes of  chiral gauge theories, the so-called  generalized  Bars-Yankielowicz (BY) and Georgi-Glashow (GG) models.

After discussing the new, generalized anomalies and their implications in various kinds of chiral gauge theories,    we  explore  in Sec.~\ref{sec:stronganomaly} the
 implications of the  strong anomaly on the phases of  the same classes of  chiral gauge theories,  studied  in Sec.~\ref{Method1}  $\sim$  Sec.~\ref{BYGG}.  

We conclude in Sec.~\ref{conclude} by summarizing the results found, and by discussing interesting analogies and contrasts between the dynamics of massless QCD 
and  chiral gauge theories. A clearer picture of  infrared dynamics of many strongly-coupled chiral gauge theories seems to emerge. 

Appendices \ref{conf1} - \ref{Higgs4} are a  collection of tables summarizing the massless fermions and their quantum numbers in various possible phases of BY and GG models.

\section{Computation of  the  mixed anomalies  \label{Method1} }

Gauging of a discrete (1-form) center symmetry and calculating anomalies induced by it in some otherwise nonanomalous global discrete symmetry  
 -  a generalized 't Hooft anomaly -   is the central theme of the work reviewed here.  Let us go through the basic elements of the analysis and enlist main formulas needed  to get to the physics results discussed in the subsequent sections.   For more general introduction and theoretical considerations on the generalized symmetries    the reader can 
 consult the original literature \cite{AhaSeiTac}-\cite{BKL4}.

 We need to distinguish two different classes of models:  the first concerns the systems where the fermions do not transform under a ${\mathbbm Z}_k$ ($k$ is a divisor of $N$) subgroup of the $SU(N)$ center. These systems possess a ${\mathbbm Z}_k$ 1-form symmetry (the "center symmetry"), which acts naturally on fundamental Wilson loops. Their analysis is  relatively straightforward.    
 In the second class of system the fundamental fermions transform non-trivially under the center of the gauge group, ${\mathbbm Z}^{\rm c}_N$, and only the diagonal combination ${\mathbbm Z}_N \subset {\mathbbm Z}^{\rm c}_N\times G_{\rm f}$ (being $G_{\rm f}$ the flavor symmetry group) leaves them invariant. The study of these models requires a careful determination of the global structure of the symmetry group involved.

\subsection{Gauging a 1-form ${\mathbbm Z}_k \subset  {\mathbbm Z}_N$   center symmetry  \label{review}}
 
First consider  $SU(N)$ theories  with an exact center ${\mathbbm Z}_k \subset  {\mathbbm Z}_N$ symmetry, $k$ being a divisor of $N$, 
{\it   under which the matter fermions do not  transform}.   Examples are: pure $SU(N)$ YM theory or the adjoint  QCD,  where $k=N$,  or various models with fermions neutral with respect to  some  ${\mathbbm Z}_k$, see Sec.~\ref{Simplecases}.

The procedure was  formulated in \cite{GKSW}    building upon some earlier results \cite{Seiberg}-\cite{AhaSeiTac},     and used in \cite{GKKS} for $SU(N)$ Yang-Mills theory at $\theta=\pi$.   The methods have been further developed and  found other areas of applications
  \cite{ShiYon}-\cite{Anber:2019nfu}.   

1-form center symmetry can be simply understood in the formalism of principal bundles. Here the gauge and the fermions fields are defined locally on open patches $U_i$ of our spacetime. These local definitions are glued together by $SU(N)$ valued transitions functions, $g_{ij}: U_i \cap U_j \rightarrow SU(N)$. In particular,
\be
\psi_i(x) = R(g_{ij}(x))\psi_j(x) \quad x \in U_i \cap U_j\;, \label{eq:fcc}
\ee
where $\psi_i$ and $\psi_j$ are the local expressions of the field $\psi$ (which transform in the representation R) in the patches $U_i$ and $U_j$.

To require that the theory is an $SU(N)$ theory (i.e. the fundamental Wilson loops are meaningful) enforces the cocycle condition, 
\be
g_{ij}g_{jk}g_{ki}=\mathbbm 1\;, \label{eq:coc1}
\ee 
in the triple intersection, $U_{ijk}=U_i \cap U_j \cap U_k$.

In this language a {\it global 1-form symmetry} transformation multiplies the transition functions $g_{ij}$ by $\mathbbm Z_k$ elements, $z_{ij}$ (one for each simple intersection, $U_{ij}=U_i \cap U_j$), which satisfy their own consistency condition
\be
z_{ij}z_{jk}z_{ki}=1\;.
\ee
This transformation is a symmetry of the system if it does not spoil the equation (\ref{eq:fcc}), i.e. if $\mathbbm Z_k$ does not act on fermions. However, it can act non-trivially on fundamental Wilson loops.\footnote{It acts on non-contractible Wilson loop, therefore global 1-form gauge transformations are indexed by elements of $H^1(\mathcal M, \mathbbm Z_k)$. The $z_{ij}$ implement a 
\v{C}ech version of this cohomology group.}

If one relaxes the cocycle consistency condition, allowing 
\be
z_{ij}z_{jk}z_{ki}=z_{ijk} \in \mathbbm Z_N\;,
\ee
one obtains a {\it gauge 1-form symmetry} transformation. In this case the condition (\ref{eq:coc1}) does not make sense, and must be replaced by
\be
g_{ij}g_{jk}g_{ki}=B_{ijk} \in \mathbbm Z_k\;,
\ee
where the new data, $B_{ijk}$, are (a discretized version of) a 2-form connection.\footnote{Similarly, the $B_{ijk}$ are representatives of the second \v{C}eck cohomology group, $H^2(M, \mathbbm Z_k)$. The closeness of $B$ can be seen on quadruple overlaps.} This construction defines an $\frac{SU(N)}{\mathbbm Z_k}$ gauge bundle. If one consider the $B_{ijk}$ data dynamical, summing on them in the functional integral, one obtains a $\frac{SU(N)}{\mathbbm Z_k}$ gauge theory.

In \cite{KapSei} and \cite{GKSW} a useful construction is presented that reproduces this gauging in terms of continuous fields. We adopt this description, which is  reviewed below briefly.

The rough idea is to replace the discrete $\mathbbm Z_k$ 1-form symmetry with a $U(1)$ 1-form symmetry, at the price of introducing other new degrees of freedom. Gauge fixing these new degrees of freedom, one can gauge-fix most of the continuous 1-form symmetry. What remains is the discrete 1-form symmetry.
 
 As a first step, one must introduce a pair of $U(1)$ 2-form and 1-form\footnote{In most part of this review a compact differential-form notation is used.  For instance, 
 	$a \equiv  T^{\rm c}  A_{\mu}^{\rm c}(x)   \,dx^{\mu}$;    $F= d a + a^2 $\;;
 	$F^2 \equiv  F \wedge F =   \frac{1}{2}  F^{\mu \nu} F^{\rho \sigma} dx_{\mu} dx_{\nu} dx_{\rho} dx_{\sigma}= \frac{1}{2}
 	\epsilon_{\mu \nu \rho \sigma}   F^{\mu \nu} F^{\rho \sigma}  d^4x =   F^{\mu \nu} {\tilde F}_{\mu \nu}  d^4x$\,, and so on. }   
   gauge fields $\big(B_{\rm c}^{(2)},B_{\rm c}^{(1)}\big)$  such that 
   \cite{GKSW}
\be   {k}  B^{(2)}_\rmc =  d B^{(1)}_\rmc\;,  \label{constr}
\ee
satisfying
\beq
  B^{(2)}_\rmc \to B^{(2)}_\rmc+d \lambda_\rmc\;,\qquad B^{(1)}_\rmc \to B^{(1)}_\rmc+{k}\lambda_\rmc\;.     \label{Zn2}
\eeq
$\lambda_\rmc$ is  a 1-form gauge function.
The $SU(N)$ gauge field $a$ is embedded into a  $U(N)$ field,
\be
\widetilde{a}=a+\frac{1}{k}B^{(1)}_\rmc, 
\ee
and one requires  invariance under $U(N)$ gauge transformation. The gauge field tensor $F(a)$ is replaced by 
\be   F(a) \to   {\tilde F}({\tilde a})  -  B^{(2)}_\rmc    \;.  
\ee
  This fixes  the manner
 these $\mathbb{Z}_k^{\rm c}$ gauge fields are coupled  to the standard gauge fields  $a$.    The matter  fields must also be 
 coupled to the $U(N)$  gauge fields, so that the 
 1-form gauge invariance, (\ref{Zn2})   is  respected.  For a Weyl fermion $\psi$ in the representation  $R$  this can be done having the kinetic term as 
\be     {\bar \psi}   \, \gamma^{\mu}    \left( \partial  +  R({\tilde a}) -  \frac {{\cal N}(R)}{k}  B_{\rm c}^{(1)} \right)_{\mu}  P_L   \, \psi  \;, 
\ee
where $R({\tilde a}) $ is the appropriate  matrix form  for the representation; ${\cal N}(R)$ is the $N$-ality of $R$.   $P_L$ is the projection operator on the left-handed 
fermions.

We  introduce an external $U(1)_{\psi}$ gauge field $A_{\psi}$ to study the anomaly, e.g.,   of  a   $U(1)_{\psi}$
symmetry  $\psi  \to   e^{ i \alpha}  \psi$,   or  of a discrete subgroup of it, 
and couple it to the fermion as 
\be     {\bar \psi}   \, \gamma^{\mu}    \left( \partial  +  R({\tilde a})  -  \frac {{\cal N}(R)}{k}  B_{\rm c}^{(1)}   +  A_{\psi}  \right)_{\mu}  P_L   \, \psi  \;.\label{from}
\ee
The standard anomaly calculation for $\psi  \to   e^{ i \alpha}  \psi \simeq  \psi +i \alpha  \psi  $,    gives
\be  \delta  S_ { \delta A_{\psi}^{(0)} }  =     \frac {2\,  T(R) }{ 8\pi^2}   \int   {\tr}  F^2   \,    \delta \alpha    =    2  \, T(R)  \,     {\mathbbm Z} \,   \delta   \alpha\;. 
\ee
$ {\mathbbm Z} $ is the integer  instanton number, and it leads to  the well-known result  that a discrete subgroup 
\be     {\mathbbm Z}_{2  T(R)} \subset U(1)_{\psi}
\ee
remains. $T(R)$  is twice the Dynkin index,  
\be     \tr\, T^a T^b  =\delta^{ab}    D(R) \;, \qquad       D\Big( \raisebox{-3pt}{\yng(1)}\Big)  = \frac{1}{2}\;,     \qquad   T(R) \equiv   2\, D(R)\;.\label{DynkinIn} 
\ee
With $\big(B_{\rm c}^{(2)}, B_{\rm c}^{(1)}\big)$ fields   in Eq.~(\ref{from}),  $ U(1)_{\psi}$ symmetry can further  be broken
 due to the replacement, 
\be     {\tr}   F^2  \to     {\tr} \big( {\tilde F}-B^{(2)}_{\rm c}\big)^2\;.   \label{replace} 
\ee
Indeed,
\be  \frac{1}{8\pi^2}  \int_{\Sigma_4}     {\tr}     \big( {\tilde F}-B^{(2)}_{\rm c}\big)^2 =  \frac{1}{8\pi^2}   \int_{\Sigma_4}    \big\{   {\tr}   {\tilde F}^2 -  N    (B^{(2)}_{\rm c})^2 \big\} \;:  \label{firstof}
\ee
we recall that   $B^{(2)}_{\rm c}$  is Abelian, $\propto {\mathbbm 1}_N$, and that   $  {\tr}   {\tilde F} = N    \, B^{(2)}_{\rm c}$.
The first term  is an integer.
 The second is 
\be   -     \frac{N}{8\pi^2}    \int_{\Sigma_4} \big( B^{(2)}_{\rm c}\big)^2  =   -     \frac{N}{8\pi^2  k^2}    \int_{\Sigma_4}  dB^{(1)}_{\rm c}  \wedge dB^{(1)}_{\rm c}       = \frac{N}{k^2}\, {\mathbbm Z}\;,    \label{usethis}
\ee
which is generally  fractional. This explains  the origin of various  0-form-1-form (mixed) anomalies and the consequent stronger anomaly conditions  in many models discussed in  Sec.~\ref{Simplecases}.

\subsection{Color-flavor locked  ${\mathbbm Z}_N$   center symmetry: Master formula \label{review2}}

Subtler situations present themselves, when a gauge theory of our interest  contains  matter Weyl fermions in the fundamental,  or antifundamental,  representation of the gauge group $SU(N)$.   Ordinarily, this means that the center symmetry is lost,   leaving no possibilities of gauging the 1-form  ${\mathbbm Z}_N$  center symmetry.  Actually,  in order to consider the 't Hooft anomalies one must externally gauge also the flavor symmetry group, $G_{\rm f}$.  Having done so, in the systems of our  interest $SU(N)_{\rm c} \times G_{\rm f}$ is found not to act faithfully.    In particular, there is a $\mathbbm Z_N$ subgroup that leaves all the fields invariant. In other words,  there is a $\mathbbm Z_N$ ''color-flavor-locked" 1-form symmetry.\footnote {This hinges upon a quite remarkable property of the generalized symmetries,  
that they are all Abelian. This reflects the fact it is not possible to define time ordering between two extended operators, hence impossible to define equal-time commutators between them.  In the case of particles,   how the (equal-time) commutators can arise in the operator formalism, as a  limit of time-ordered products taken in different orders,  is best explained in Feynman's book on Path-Integral formulation of quantum mechanics \cite{Feynman}.}
Similarly to the previous case, gauging of  this 1-form symmetry allows  us to  gauge   the faithful symmetry group of the system, $\frac{SU(N)_{\rm c}\times G_{\rm f}}{\mathbbm Z_N}$.\footnote{One should keep in mind that there are "more"  configurations  in  $\frac{SU(N)}{\mathbbm Z_k}$ ($\frac{SU(N)_{\rm c} \times G_{\rm f}}{\mathbbm Z_N}$) than  in the  $SU(N)$ ($SU(N)\times G_{\rm f}$) gauge bundles, i.e. any of the latter  always belong to the former, but not the other way around.}

To introduce this kind of systems, and to discuss the method of analysis developed,   we consider the concrete example of  
an  $SU(N)$ gauge theory  with matter left-handed fermions in the reducible, complex representation, 
\be       \yng(2) \oplus   (N+4) \,{\bar   {{\yng(1)}}}\,    \label{either1}   \ee
that is, 
\beq
   \psi^{\{ij\}}\,, \quad    \eta_i^B\, , \qquad  {\footnotesize  i,j = 1,2,\ldots, N\;,\quad B =1,2,\ldots , N+4}\;,    \label{either111} 
\eeq
(which is the simplest of the so-called  Bars-Yankielowicz models).   This model will be referred to  as  the ''$\psi\eta$"  model below.

The symmetry group of the model  is 
\be   G_{\rm f}=   SU(N+4) \times U(1)_{\psi\eta}\;,     \label{beyond1}
\ee
where $U(1)_{\psi\eta}$ indicates the anomaly-free combination of $U(1)_{\psi}$ and  $U(1)_{\eta}$, associated with the two types of  matter Weyl fermions of the theory.  In this model, 
 the conventional 't Hooft anomaly matching discussion allows, apparently,  a confining phase, with no condensates and with full unbroken global symmetry, and with some simple set of massless composite fermions - ''baryons" -  saturating 
the anomaly matching equations, see Appendix~\ref{conf1}.   Notably,   the anomaly constraints are  {\it  also}    consistent with a dynamical Higgs phase, in which the color and (part of) the flavor symmetry are dynamically broken by certain bi-fermion condensates, see Appendix~\ref{Higgs1}.

The situation is analogous in a large class of chiral gauge theories, to be discussed in Sec.~\ref{BYGG} below. Clearly, the conventional 't Hooft anomaly matching requirement is not powerful enough to discriminate among possible (confining or dynamical Higgs) vacua.

To go beyond the conventional (perturbative) 't Hooft anomaly analyses,   it is necessary  to consider the global properties of the symmetry groups, not only the algebra.      For even $N$ the true  symmetry group of the  model  is found to be \cite{BKL2}: 
 \be
SU(N)_{\rm color} \times G_{\rm f}\;, \qquad  G_{\rm f}=\frac{SU(N+4) \times  U(1)_{\psi\eta}  \times (\mathbb{Z}_2)_F } {\mathbb{Z}_{N}\times \mathbb{Z}_{N+4}}\;,
\label{symmetryNevenpsi}
\ee
and not  (\ref{beyond1}),  
where  $(\mathbb{Z}_2)_F$  is the fermion parity,    $\psi, \eta \to  -\psi, -\eta$. 

Indeed, as promised, there is a subgroup of $SU(N)_{\rm c}\times SU(N+4)\times U(1)_{\psi\eta}\times (\mathbbm Z_2)_F$, 
\be      {\mathbbm Z}_N = SU(N) \cap  \{  U(1)_{\psi\eta}  \times   (\mathbb{Z}_2)_F \}\;,   \label{colorflavor}
\ee
which leaves the matter fields invariant.\footnote {There is another independent subgroup, $\mathbbm Z_{N+4}$, which does not act on matter filed, leading to another $\mathbbm Z_{N+4}$ 1-form center symmetry. In \cite{BKL2}  the effects of gauging  this flavor center symmetry and the resulting mixed anomalies in the $\psi\eta$ model have also been taken into account.  None of the main results 
	however were found to depend on it.  Here for simplicity we consider only the gauging of the color-flavor locked center symmetry ${\mathbbm Z}_N$, together with
	$U(1)_{\psi\eta}$ and $({\mathbbm Z}_2)_F$.}  The gauge transformation with $\rme^{\frac{2\pi\im}{N}}\in \mathbb{Z}_N\subset SU(N)$, 
\be
\psi\to \rme^{\frac{4\pi\im}{N}}\psi\;,\; \qquad \eta \to \rme^{-\frac{2\pi\im}{N}}\eta\;,    \label{ZNpsieta}
\ee
can be undone by the following $(\mathbb{Z}_2)_F \times  U(1)_{\psi\eta}  $ transformation:
\be
\psi \to (-1)\, \rme^{\im\frac {N+4}{ 2}\frac {2\pi}{ N}}\psi = \rme^{-\im \frac{N}{2}\frac {2\pi}{ N}}   \, \rme^{\im\frac {N+4}{ 2}\frac {2\pi}{ N}}\psi \;, \qquad \eta\to (-1)\, \rme^{-\im \frac{N+2}{2}\frac {2\pi}{ N}}\eta=    \rme^{\im \frac{N}{2}\frac {2\pi}{ N}}     \, \rme^{-\im \frac{N+2}{ 2}\frac {2\pi}{ N}}\eta\;. \label{ZNequiv}
\ee
A relevant fact is  that the odd elements of $\mathbb{Z}_N$ belong to the disconnected component of $U(1)_{\psi\eta}\times (\mathbb{Z}_2)_F$ 
whereas   the even elements belong to the connected component of the  identity. 

The presence of a subgroup which acts trivially means that there is a 1-form global symmetry. Again, in the discrete language introduced before, it acts on transition functions. In particular, if $g^{\rm c}_{ij}$, $u_{ij}$ and $q_{ij}$ are the transition functions for $SU(N)$, $U(1)_{\psi\eta}$ and $ (\mathbbm Z_2)_F$, one may introduce some $\mathbbm Z_N$ transitions functions (a $\mathbbm Z_N$ gauge field), $z_{ij}$, and transform
\be
g_{ij}\rightarrow z_{ij}g_{ij}\;, \quad u_{ij}\rightarrow (z_{ij})^{-1} u_{ij}\;, \quad\text{and}\quad  q_{ij} \rightarrow (z_{ij})^{-\frac{N}{2}}\;q_{ij}\;.\label{eq:cfl_1faction}
\ee  
If one drops the cocycle condition for $z_{ij}$, one gauges the 1-form symmetry. In this case one must introduce also the 2-form connection \footnote{Again, an element of $H^2(M, \mathbbm Z_N)$, $B_{ijk}\in \mathbbm Z_N$.}, described by the new data $B_{ijk} \in \mathbbm Z_N$, which are read from the transition functions
\be
g_{ij}g_{jk}g_{ki}=B_{ijk}\;, \quad u_{ij}u_{jk}u_{ki}=(B_{ijk})^{-1}\;, \quad q_{ij}q_{ji}q_{ki}=(B_{ijk})^{-\frac{N}{2}}\;.
\ee
This definition assures that all fields  (matter, gauge) are well defined in the triple intersections.
 
Again, let us turn to the continuous language.
As a first step, we have to gauge $U(1)_{\psi\eta}\times  (\mathbbm Z_2)_F$, introducing
\begin{enumerate}
\item $A$: $U(1)_{\psi\eta}$ 1-form gauge field, 
\item $A_2^{(1)}$:  $({\mathbbm Z}_{2})_{F}$  1-form gauge field,
\end{enumerate}
in addition to the dynamical color gauge $SU(N)$ field,  $a$.
  
  The gauging of 1-form  discrete  $\mathbb{Z}_{N}$   symmetry
is done by  introducing 
\be  N  B_\rmc^{(2)} = d  B_\rmc^{(1)}\;.   \label{1fgconstraint}
\ee
These 2-form gauge fields must be coupled to the  $SU(N)$  gauge fields  $a$ and $U(1)_{\psi\eta}  \times  (\mathbb{Z}_2)_F$ gauge fields   ($A$ and $ A_2^{(1)}$)    appropriately.  
We first embed $a$ in  a  $U(N)$ gauge field $\widetilde{a}$ as 
\be
\widetilde{a}=a+\frac{1}{N}B^{(1)}_\rmc
\ee
and  requiring the  invariance 
\begin{align}  B_\rmc^{(2)} & \to B_\rmc^{(2)}+\diff \lambda_\rmc\;, \qquad
 B_\rmc^{(1)}  \to B_\rmc^{(1)}+ N  \lambda_\rmc    \;,   \nonumber \\ 
 \widetilde{a} &\to \widetilde{a}+\lambda_\rmc\;.
   \label{werequire} 
\end{align}
 In these equations $\lambda_\rmc$ is a properly normalized $U(1)$ gauge field, which satisfies its own Dirac quantization condition.

To reproduce (\ref{eq:cfl_1faction})  correctly in this continuous language, $U(1)_{\psi\eta} $ and  $  (\mathbb{Z}_2)_F$ gauge fields must also transform,   
\be  A\to A-\lambda_\rmc\;, \qquad    A_2^{(1)}\to A_{2}^{(1)}+\frac{N}{ 2}\lambda_\rmc   \;.   \label{simult}\;
\ee
The last equation needs a comment, as $A^{(1)}_2$ is a $\mathbbm Z_2$ gauge field, while $\lambda_\rmc$ is a $U(1)$ gauge field. To be precise, it is more correct to proceed as it has been done with the $SU(N)$ gauge field. In particular one should write a $U(1)$ gauge connection
	\be
	\tilde A_2 = A_2 + \frac{1}{2} B^{(1)}_c, \label{eq:a2def}
	\ee
	and impose
	\be
	\tilde A_2 \rightarrow \tilde A_2 + \frac{N}{2} \lambda_\rmc\;.
	\ee
	As before $a$ is not a globally defined $SU(N)$ gauge field while $\tilde a$ is a correctly normalized $U(N)$ gauge field, and  now $\tilde A_2$ is a correctly normalized $U(1)$ field.

One has now an  $\frac{SU(N)}{{\mathbbm Z}_N}$ connection rather than  $SU(N)$. It implies that 
 \be      \frac{1}{2\pi} \int_{\Sigma_2}    B_\rmc^{(2)}    =     \frac{ n_1 }{N}\;,    \qquad     n_1 \in   {\mathbbm Z}_N\;,   \label{Byconstruction2}
 \ee
  in a closed two-dimensional subspace,    ${\Sigma_2}$.    On a  topologically nontrivial four dimensional spacetime
 of Euclidean signature which contains such sub-spaces one has 
  \be      \frac{1}{8\pi^2} \int_{\Sigma_4}   (B_\rmc^{(2)})^2   =   \frac{n }{N^2}\;,\label{Byconstruction4}
 \ee
   where $n \in   {\mathbbm Z}_N$.
  
     The fermion kinetic term with the background gauge fields is determined by the minimal coupling procedure as 
\bea
&&\overline{\psi}\gamma^{\mu}\left(\partial +\calR_{\rmS}(\widetilde{a})+\frac{N+4}{ 2}A+\tilde A_2\right)_{\mu}P_\rmL\psi\;  \nonumber\\
&&+\,\overline{\eta}\gamma^{\mu}\left(\partial +\calR_{\rmF^*}(\widetilde{a}) -\frac{N+2}{2}A-\tilde A_2\right)_{\mu}P_\rmL\eta\;. \label{naive}
\eea
(with an obvious notation).   Note that each of the kinetic terms  is  invariant under  (\ref{werequire}) and (\ref{simult}).

The 1-form gauge invariance of our system  can be made completely  manifest, by rewriting the above as
\bea
&&\overline{\psi}\gamma^{\mu}\left(\partial +  [\calR_{\rmS}(\widetilde{a})- \frac{2}{N} B_\rmc^{(1)}]     +\frac{N+4 }{ 2}  [A + \frac {1}{N} B_\rmc^{(1)}]     + [\tilde A_2-   \frac{1}{2}  B_\rmc^{(1)} ]  \right)_{\mu}P_\rmL\psi\;  \nonumber\\
&&+\,\overline{\eta}\gamma^{\mu}\left(\partial  +[\calR_{\rmF^*}(\widetilde{a}) +\frac{1}{N} B_\rmc^{(1)} ] -\frac{N+2}{ 2} [A + \frac {1}{N} B_\rmc^{(1)}]   - [\tilde A_2-   \frac{1}{2}  B_\rmc^{(1)}]   \right)_{\mu}P_\rmL\eta\;.    \nonumber \\  \label{notnaive}
\eea
Written this way, the expression inside each square bracket is  invariant under   (\ref{werequire}) and (\ref{simult}).  
This leads to the gauge field strength for the $\psi$ and $\eta$, in the form used in the analysis \`a la  Stora-Zumino descent procedure \cite{Stora,Zumino}, in \cite{BKL2,BKL4}.  
The final answer of course does not depend on the rewriting of the kinetic terms as  (\ref{notnaive});  the original form (\ref{naive})  is perfectly adequate for the calculation of the anomaly  in a more straightforward approach explained below.

 Under the fermion parity, such as  $\psi \to - \psi$, $\eta \to - \eta$  in the $\psi \eta$ model,   the contribution to the  $(\mathbbm Z_2)_F - [{\mathbbm Z}_N]^2$  anomaly from a fermion in the representation ${R}$  is  given by the phase  in the partition function,  
 \be     c_2  \,  \frac{1}{8\pi^2}\int_{\Sigma_4}   {\tr}_{R}   \left[\big(F(\tilde{a})\big)^2\right]   \, (\pm \pi)  
 \ee
 where $c_2$  is the ${\mathbbm Z}_2$  charge of the fermion. 
   In the case of the $\psi\eta$ model, 
 for instance,  $c_2(\psi)=1$,  $c_2(\eta)=-1$, see (\ref{naive}).

  Now
\bea  
{\tr}_{R}   \left[\big(F(\tilde{a})\big)^2\right] &=&  {\tr}_{R}\left[\big(F(\tilde{a}) - B^{(2)}_{\rm c} + B^{(2)}_{\rm c}\big)^2\right]    \nonumber\\
&=& {\tr} \left[ \left(\mathcal{R}_R     \big( F(\tilde{a}) - B^{(2)}_{\rm c}  \big)     +   {\cal N}(R) B^{(2)}_{\rm c}    {\mathbbm 1}_{d(R)}   \right)^2    \right]  \nonumber\\
&=&{\tr}   \left[ \mathcal{R}_R     \big(F(\tilde{a}) - B^{(2)}_{\rm c})^2   + {\cal N}(R)^2    \big(B^{(2)}_{\rm c} \big)^2   {\mathbbm 1}_{d(R)}     \right]  \;.
\eea
 $\mathcal{R}_R$ is the  matrix form for the representation $R$ and   ${\cal N}(R)$ its $N$-ality,    and we used  the fact that 
\be     {\tr}_{R}   \big(F(\tilde{a}) - B^{(2)}_{\rm c} \big)  = 0\;,
\ee
valid for an $SU(N)$  element in a general  representation.  $ {\mathbbm 1}_{d(R)} $ is  the  $d(R)\times d(R)$   unit matrix   ($d(R)$ is the dimension of the representation $R$).  
One finds
\bea
 {\tr}_{R}   \left[\big(F(\tilde{a})\big)^2\right] &=&   D(R)\,  {\tr}_{F}\left[\big(F(\tilde{a}) - B^{(2)}_{\rm c}\big)^2\right] + d(R)  {\cal N}(R)^2\big(B^{(2)}_{\rm c}\big)^2 =\nonumber\\
&=& D(R)\,   {\tr}_{F}\left[F({\tilde a}) \right]^2 + \left[  - D(R) \cdot N +   d(R)   {\cal N}(R)^2  \right]  \big(B^{(2)}_{\rm c}\big)^2\;,  \label{result}
\eea
where  $D(R)$ is twice  the Dynkin index  $T_R$,    (\ref{DynkinIn}).
Note that
\be      \frac{1}{8\pi^2}\int_{\Sigma_4}     {\tr}_{F}\left[F({\tilde a})^2 \right]   \in {\mathbbm Z}\;:  
\ee
 the first term in Eq.~(\ref{result})  corresponds to the conventional instanton contribution to the  $({\mathbbm Z}_2)_F$ anomaly.  
 In all models of interest here, however,   the {\it  sum} of the instanton contribution from the fermions is  of th form, 
 \be          ({\rm an} \, {\rm even}\, {\rm integer }) \times   \frac{1}{8\pi^2}\int_{\Sigma_4}   {\tr}_{F}\left[F({\tilde a})^2 \right] \, \times ({\pm \pi})   =  2\pi {\mathbbm Z}\;,
 \ee  
 which is trivial. 
 
 {The fact that   $({\mathbbm Z}_2)_F$ anomaly  is absent in the standard instanton analysis  because of a (nonvanishing)  {\it even} coefficient, and of the quantized
 instanton flux, but {\it  not} because of an algebraic cancellation from different fermions, {\it   is of utmost importance}.   Indeed, the gauging of the 1-form ${\mathbbm Z}_N$ by the introduction of the 2-form gauge fields  $B^{(2)}_{\rm c}$  basically  amounts to  the fractionalization of the instanton flux \`a la 't Hooft,  (\ref{Byconstruction4}), and as a consequence, a nonvanishing mixed anomaly involving $({\mathbbm Z}_2)_F$ can appear.

Thus   the non-vanishing  mixed  $({\mathbbm Z}_2)_F-  [{\mathbbm Z}_N]^2$  anomaly comes only from  the second term of  Eq.~(\ref{result}),
containing the 2-form gauge field,   
\be    \boxed{      \Delta S^{({\rm Mixed}\, {\rm anomaly})}  =  (\pm \pi) \cdot  \sum_{\rm fermions}   c_2  \,  \Big(d(R)  {\cal N}(R)^2- N \cdot D(R)\Big) \, \frac{1}{8\pi^2}\int_{\Sigma_4}       \big(B^{(2)}_{\rm c}\big)^2 \;.    \label{masterF}   }
\ee
This is the master formula.

The  formula (\ref{masterF})  gives  the result for the mixed anomaly in {\it all} models considered in   \cite{BKL2, BKL4} at once.
 For instance, for the  $\psi\eta$ model, one gets 
 \bea   &&   \Delta S^{({\rm Mixed}\, {\rm anomaly})}   \nonumber   \\  
 &=&      \frac{\pm\pi}{N^2} \left[  \left(\frac{N(N+1)}{2}\cdot 4   - N(N+2)\right)
 -  (N+4)  \left(    N \cdot 1 -   N \cdot 1  \right) \right]     =\pm\pi \;,  \label{psietaOK}
\eea
which means that  there is a  $({\mathbbm Z}_2)_F$ anomaly in the presence of   the $ {\mathbbm Z}_N$ gauging.  More precisely,    there is an obstruction for gauging simultaneously 
  ${\mathbbm Z}_N$,  
$U(1)_{\psi\eta}$  and $(\mathbb{Z}_2)_F $,   by keeping the  equivalence
\be      {\mathbbm Z}_N \subset  SU(N) \sim       {\mathbbm Z}_N  \subset  \{  U(1)_{\psi\eta}  \times   (\mathbb{Z}_2)_F \}\;.   \label{colorflavorBis}
\ee

 One can repeat the same calculation in the possible confining phase without symmetry breaking,  Appendix~\ref{conf1}. The result is very simple: there is no such anomaly. This can be traced back to the fact that the massless baryons are $SU(N)_{\rm c}$ singlet, and any appearance of $B^{(1)}_{\rm c}$ in their covariant derivative simply cancel.
Clearly this mismatch of anomaly {\it forbids confinement without symmetry breaking}.

The same cannot be said in the dynamical Higgs scenario, see Appendix~\ref{Higgs1}, as the color group is broken.

Even though, for concreteness,  we discussed above the particularly simple model, the $\psi\eta$ model,  
the master formula found above is actually applicable to any theory,   after the correct symmetry is found and  after the  fermion kinetic terms, invariant under the  1-form  $ {\mathbbm Z}_N$  gauge symmetry  are written down. 
The results for the    $({\mathbbm Z}_2)_F-  [{\mathbbm Z}_N]^2$  anomaly 
in   the $\chi\eta$ model as well as   all other   
generalized  Bars-Yankielowicz and 
Georgi-Glashow models   \cite{BKL2}, \cite{BKL4}  discussed below in Sec.~\ref{BYGG}  indeed  follow straightforwardly from the master formula   (\ref{masterF}) this way.   See Sec.~\ref{BYGG} below.    
 
%
%

\subsection{Comments on the paper \cite{Tong}}

In a recent paper \cite{Tong} the $\psi\eta$  model and $\chi\eta$ model (in our notation) are studied, and the authors claim 
that  ``there is no  $({\mathbbm Z}_2)_F$ anomaly",   of the type discussed in the previous section. 
Such a statement is however intrinsically  ambiguous.  It is unclear whether the authors'  claim is that there is no  $({\mathbbm Z}_2)_F$ symmetry,  or that there is 
one but is nonanomalous.  

Indeed, their  argument in their Sec. 2.1 seems to indicate the former; but as we have made explicit here  in (\ref{symmetryNevenpsi}) \cite{BKL4}, an independent 
  $({\mathbbm Z}_2)_F$ symmetry exists, but only in the $SU(N)/{\mathbbm Z}_N$  gauge theory, not in the original $SU(N)$ theory.    
   Their  comments in Sec.~2.5  also seem to be in line with  the first.   But the fact  that  the $({\mathbbm Z}_2)_F$ symmetry we are interested in here coincides with the angle $2\pi$ space rotation, is well known, and it has been taken into account in our papers \cite{BKL2,BKL4}. Any $({\mathbbm Z}_2)_F$ anomaly could be cancelled by a space rotation, so in that sense, there would never be a  $({\mathbbm Z}_2)_F$ anomaly.  But this is not the point.   As there is no a priori guarantee that  the Lorentz invariance cannot be dynamically broken,  a $({\mathbbm Z}_2)_F$ anomaly arising from the gauge dynamics  cannot be allowed,  if the Lorentz invariance is to be maintained. 
   
   Their discussion in Sec.~3  about the Higgs phase in these models does not contain anything new, as compared to what we have discussed about the Higgs phase, see \cite{BKL4},   and    Sec.~\ref{higgs_remark} and  in Appendices  \ref{Higgs1}, \ref{Higgs2}, \ref{Higgs3}, \ref{Higgs4}, in this work,    for the $\chi\eta$, $\psi\eta$  models and for all other  BY and GG models.  
  
The main point of  the paper   \cite{Tong} seems to be in  Sec. 2.2,   which apparently leads to the second conclusion, that  there is a  $({\mathbbm Z}_2)_F$ symmetry but is non anomalous. 
  They argue that,    by choosing the normalization of the $\mathbbm Z_2$ gauge field  as 
\be     \int_\Sigma  dA_2^{(1)}    =  2\pi \,{\mathbbm Z}\;,   \label{Norm2} 
\ee
(a formula in the line below Eq. (2.13)  of     \cite{Tong}),
the relation  
 \be       2   A_2^{(1)} -   B^{(1)}_\rmc -  B^{(1)}_\rmf     =    d   A_2^{(0)}\;,    \label{both}
\ee
leads to 
\be     2   \,   \diff   A_2^{(1)}-{N}B^{(2)}_\rmc-  (N+4) B^{(2)}_\rmf  =0 \;:   \label{this}
\ee
by taking the derivatives of the both sides, hence to the constraints  on the fluxes of  $\mathbbm Z_2$  and  $\mathbbm Z_{N+4}$  gauge fields,
\be    \int_{\Sigma_2} N\, B^{(2)}_\rmc  +    \int_{\Sigma_2}  (N+4) \, B^{(2)}_\rmf  =      4   \pi k\;,\qquad  k \in {\mathbbm Z}\;.  \label{consistentwith}
\ee
If one chooses not to  introduce  the 1-form gauging   $ B^{(2)}_\rmf  $  (as we did in \cite{BKL4})  one would simply  get 
\be    \int_{\Sigma_2} N\, B^{(2)}_\rmc  =      4  \pi k\;,\qquad  k \in {\mathbbm Z}\;,  \label{Noresult}
\ee
and  our anomaly (\ref{masterF})  would indeed disappear.     The rest of Sec. 2 in  \cite{Tong}   all follows from the normalization,  (\ref{Norm2}).

However,  (\ref{Norm2})   means that their background  $\mathbbm Z_2$  gauge field corresponds to 
\be   \psi \to \psi\;, \qquad  \eta \to \eta\;, 
\ee
i.e.,   no transformation  (the trivial element of $\mathbbm Z_2$).   The fact that one finds no anomaly in such a background is  certainly correct, but it is not what one is interested in.

The correct normalization for a  $\mathbbm Z_2$  gauge field is  the one we have adopted,  
\be   \oint   A_2^{(1)}  = \frac{ 2\pi m }{2} \;,   \qquad     m\in {\mathbbm Z}\;,    \label{Z2gauge}
\ee
that (for the nontrivial element)  corresponds to the holonomy,
\be   \psi \to  - \psi\;, \qquad  \eta \to  -   \eta\;, 
\ee
This leads to  (we ignore  the ${\mathbbm Z}_{N+4}$ gauge field) 
\be   \oint dx^{\mu}    \big( 2   A_2^{(1)} -   B^{(1)}_\rmc  \big)_{\mu}     =   \oint    d   A_2^{(0)}  =   2\pi n\;,   \qquad  n\in {\mathbbm Z}\;, 
\ee
and
\be    \int_{\Sigma_2} N\, B^{(2)}_\rmc  =      2\pi k\;,\qquad  k \in {\mathbbm Z}\;,  \label{consistentwith}
\ee
and this leads to the  anomaly,   (\ref{masterF}).

In other words, our assumption is 
 that it is possible to choose the smallest cycle of $B^{(2)}$ compatible with the Dirac flux quantization for $B^{(1)}$, i.e. 
\be 
\int_\Sigma B^{(2)}=\frac{1}{N}\int_\Sigma dB^{(1)} = \frac{2\pi}{N}\;,\label{eq:flux_quantization0}
\ee 
without any topological obstruction. In the discrete language, the analogous assumption is to be able to choose $B_{ijk}$ to be any element in  $H^2(\mathcal M, \mathbbm Z_N)$.

Actually,  if one insists on working with the theory on a smooth manifold, without any topological defect for the $\mathbbm Z_2$ gauge field  $A_2$, the assumption made above cannot be maintained, as pointed out by ourselves  \cite{BKL2}.   And this seems to be the point on which  the  authors of \cite{Tong} are
trying to make  a clean  mathematical statement.



However, $A_2$ is not a proper $(\mathbbm Z_2)_F$ gauge field, as $a$ is not an $SU(N)$ gauge field. In particular its cocycle condition in triple overlap might fail, leading to curvature-like insertions at discrete points. Moreover, the 1-form gauging of $\mathbbm Z_N$ invalidates the naive Dirac quantization condition for $A_2$, as it can be checked directly: if one separates a 2D cycle $\Sigma$ through the curve $\gamma$ in $\Sigma_1\cup \Sigma_2$, one obtains
\bea
\int_\Sigma dA_2  &=& \int_{\Sigma_1} dA_2 + \int_{\Sigma_2} dA_2= \int_\gamma (A_2)_{\Sigma_1} - \int_\gamma (A_2)_{\Sigma_2}    \nonumber    \\
   &=&  \frac{N}{2} \int_\gamma \lambda_{12} =k\pi\;,  \qquad     k \in {\mathbbm Z}\label{eq:diracderivation}
\eea
(see (\ref{simult}))
as $\lambda_{12}$ (1-form gauge transition functions) \footnote{The simple fact that the $\mathbbm Z_2$ gauge field transforms non-trivially and changes from a patch to another, means that gauging of 1-form symmetry has 
been  appropriately  implemented. Indeed, thanks to (\ref{eq:diracderivation}), a Wilson loop of the form $e^{\oint A_2}$ is not 1-form gauge invariant,  i.e.,  it is not  a proper line operator. This is where we differ from part of the analysis of \cite{Tong}. } is a $\mathbbm Z_N$ 1-form gauge field, satisfying 
\be
\oint_\gamma \lambda =\frac{2\pi}{N}\;.
\ee
This leads to the more general flux quantization (\ref{eq:flux_quantization0}), and allows to insert an odd number of flux insertion in the surface.

To recapitulate,  in half of \cite{Tong} the authors argue that there is no  $(\mathbbm Z_2)_F$ symmetry; in the other half,  they discuss the background  $(\mathbbm Z_2)_F$-${\mathbbm Z}_N$ 1-form gauge fields,  corresponding however to the  trivial element of  the 1-form $(\mathbbm Z_2)_F$ transformation, finding no anomaly.



%
%
%

 \subsection{Comments on the papers \cite{Murayama1,Murayama2}}

 Two interesting papers  appeared recently, which discuss the $\psi\eta$ and $\chi\eta$ models.  The authors of \cite{Murayama1,Murayama2}  start from the ${\cal N}=1$ supersymmetric version of the models, and introduce 
 a particular  (``anomaly-mediated") supersymmetry breaking perturbation.  In the second paper (on the $\psi\eta$ model)  this is done by making use of the known (Seiberg-) duality for this systems, at the origin of the moduli space
 of this model.   In the $\psi\eta$ model, with $SU(N)$ gauge group and 
with a global symmetry group
\be   G_{\rm f}=   SU(N+4) \times U(1)_{\psi\eta}\;,   \label{beyond1111}
\ee
the authors  claim \cite{Murayama2}   that   for $N\ge   21$  the global symmetry is broken to   $SO(N)$, with no massless composite fermions,  whereas for  $N<21$  the system 
flows into a  conformal fixed point in the IR. 
 
 For the $\chi\eta$ model, with odd $N$, they argue  \cite{Murayama1}  that the global symmetry $SU(N-4)$ is spontaneously broken to $Sp(N-5)$.   For $N$ even the unbroken symmetry is claimed to be $Sp(N-4)$.

 We shall not go into the details and merits of their analyses, but will make only a few general comments on  their use of the supersymmetric models as the starting point of the analysis.   
 First of all,  in supersymmetric version of these models, there are  often nontrivial quantum moduli space of vacua (vacuum degeneracies, or flat directions),  whereas in the nonsupersymmetric chiral models 
 we are studying here the vacuum is always  unique and strongly coupled.  It is a nontrivial question which point  in the moduli space of the supersymmetric theory  (apart from which perturbation to use)  is the correct  one  to choose,  to start the analysis \footnote{This subtle problem is discussed in \cite{Cordova} in a slightly different but  basically similar  context, of  perturbing a ${\cal N}=2$ supersymmetric model  to ${\cal N}=0$ (nonsupersymmetric)  model, in the attempt of finding out  the correct  infrared dynamics of  the non supersymmetric  $SU(2)$    theories with different number of (adjoint)  flavors. }.  
 
%
%
 
  Secondly,   all bifermion condensates such as 
 $\brc \psi \eta \ckt $  (in the $\psi\eta$ model)   and   $\brc \chi \eta \ckt $  and    $\brc \chi \chi  \ckt $ (in the $\chi\eta$ model), which  are analogue of   the quark condensate in QCD, and play the central roles in the  (candidate) Higgs vacua of these  models,  are forbidden in supersymmetric version of the models,   as can be easily proven by use of supersymmetric Ward-Takahashi identities \cite{AmatiKMV}.  In other words, these condensates are absent,  unless supersymmetry is dynamically broken, which does not occur in general, 
 supersymmetric chiral gauge theories \cite{AmatiKMV}.    Also, 
 in supersymmetric models,  the global symmetry breaking  occurs due to the condensation of scalar fields, which do not exist in nonsupersymmetric theories. 
 Because of all this, the  infrared dynamics of  supersymmetric and nonsupersymmetric  theories   are usually  very different, even though the gauge group and the global symmetry group are the same.  A strong bifermion condensates such as  $\brc \psi \eta \ckt \sim \Lambda^3$ or   $\brc \chi \eta \ckt \sim \Lambda^3 $ 
 are intrinsically nonperturbative effects.
They cannot  be found via small perturbations in a theory in which they vanish  by symmetries.  
 
The crucial question whether or not a phase transition  occurs when the supersymmetry breaking mass parameters introduced  reach some critical values, seems to be unanswered 
in \cite{Murayama1,Murayama2}.

\subsection{Higgs phase  and anomaly-matching}  \label{higgs_remark}

As said above, it is possible to satisfy the standard 't Hooft anomaly matching also in the Higgs phase,  see Appendix~\ref{Higgs1}.    Even though these results are known from the earlier work \cite{BY}-\cite{BK}  and in \cite{BKL4},  the remarkable  {\it  way} it works, as compared to the matching equations in the ''confining vacua",  is perhaps not generally known.  
 The Higgs phase of these chiral theories are, in general,  described by massless NG bosons together with  some massless fermions.   These fermions saturate the conventional 't Hooft anomaly triangles   with respect to the unbroken flavor symmetries.  The way they do is, however, quite remarkable, and in our view,  truly significant. As can be seen from Table~\ref{SimplestBis}, Table~\ref{SimplestAgain2}, and in similar 
 Tables~\ref{brsuv},   \ref{brsir},   \ref{brauv} and   \ref{brair}   for the generalized BY and GG models   (in Appendices  \ref{Higgs1}, \ref{Higgs2}, \ref{Higgs3}, \ref{Higgs4}),  the set of fermions remaining massless in UV   and  those in the IR are  {\it identical}  in   their quantum numbers, charges, and multiplicities. Therefore,  the matching of anomalies (in the unbroken global symmetries) is completely automatic, and  natural. No arithmetic 
equations need be solved.   We may further argue that this way the system ''solves"  't Hooft's
anomaly-matching conditions in the true sense.   Note that this solution (Higgs phase vacua, with given sets of condensates) is stable, in the sense that 
any extra (1-form) gauging or possible new mixed anomalies would not introduce any new constraints: the matching continues to be automatic.

\section{Physics of   models with   ${\mathbbm Z}_k \subset  {\mathbbm Z}_N$ center symmetry \label{Simplecases}  }

In this section we review the study of symmetry breaking,  implied by the   various mixed anomalies of the type,   
${\mathbbm Z}_{\ell}^{(0)}  -  \big[ {\mathbbm Z}_{k}^{(1)} \big]^2 $,   where ${\mathbbm Z}_{\ell}^{(0)} $  is some 0-form (ordinary) discrete symmetry,   and 
${\mathbbm Z}_{k}^{(1)}$ is a 1-form  symmetry, based on a subgroup,   ${\mathbbm Z}_{k}\subset {\mathbbm Z}_{N}$  of the color $SU(N)$ center. 
The method of analysis has already been explained in Sec.~\ref{review}.    
${\mathbbm Z}_{\ell}$ and  ${\mathbbm Z}_{k}$ depend on the particular model considered, but as we will see,  many 
interesting models can be analyzed this way \cite{BKL1}.  

\subsection{Self-adjoint antisymmetric tensor matter}
\label{sa}

We start with  a class of  $SU(N)$ gauge theories ($N$ even) where the matter fermions are in the $\frac N2$ rank fully antisymmetric representation.  
There is a 1-form   $ {\mathbbm Z}_{\frac{N}{2}}^{\rm c} $ center symmetry  present,  and we wish to know  if some mixed 't Hooft anomaly with the 0-form (ordinary) symmetries might arise.

\subsubsection{$SU(6)$  gauge group  \label{sec:Yamaguchi}}

Our first example is an $N=6$ theory with  $N_{\rm f}$  flavors of Weyl fermions  in  the representation
\be   {\underline{20}} \, = \, \yng(1,1,1)\ .
\ee
This  ($SU(6)$) is the simplest nontrivial case    of interest, as we will see. 
Moreover, if $N_f\leq 10$ the theory is asymptotic free, and discussions basically similar to those below can be 
worked out \cite{BKL1}.
 
 There is a $U(1)_{\psi}$ global symmetry, in all these cases,   broken by the instantons  to a global discrete  ${\mathbbm Z}_{6 N_{\rm f}}^{\psi}$
 symmetry,   which is  further broken by the 1-form gauging to  ${\mathbbm Z}_{2 N_{\rm f}}^{\psi}$.
 This last step is due to a mixed  't Hooft anomaly, that is,    there is an obstruction to gauging such a 
$ {\mathbbm Z}_{\frac{N}{2}}^{\rm c} $ discrete center symmetry,  together with  the global   ${\mathbbm Z}_{6 N_{\rm f}}^{\psi}$  symmetry.


Below, we will take $ N_{\rm f}=1$.
The  model was studied   first in \cite{Yamaguchi} and then  in \cite{BKL1}.

This model has a nonanomalous 
 ${\mathbbm Z}_6^{\psi}$ symmetry,
 \be   
{\mathbbm Z}_6^{\psi}\,:\quad  \psi \to    e^{\frac{2\pi i}{6}  j} \psi \ ,   \qquad  j=1,2,\dots, 6  \ .   \label{flavorZ6}
\ee
which is a subset of $U(1)_{\psi}$.
The system has also an exact center symmetry  acting  Wilson loops as 
\be     
 {\mathbbm Z}_3^{\rm c}  \,: \quad  e^{ i \oint A } \to         e^{\frac{2\pi i}{6} k}  e^{ i \oint A }   \ ,      \qquad  k=2,4,6 \ ,
\ee
which  does not act on $\psi$.

By introducing the   $ {\mathbbm Z}_3^{\rm c}$  gauge fields,   
\be   {3}  B^{(2)}_\rmc =  d B^{(1)}_\rmc\;,  \label{constr11}
\ee
use of (\ref{replace})-(\ref{usethis})  gives,    for 
the  anomalous  ${\mathbbm Z}_6^{\psi}$  transformation,  
\be    \left ( \frac{6 }{ 8\pi^2}  \int   {\tr}  {\tilde F}^2    -  \frac{ 6 N }{ 8\pi^2}  \int (B_{\rm c}^{(2)})^2   \right)   \delta   A_{\psi}^{(0)}  \;,    \label{formula1} 
\qquad  
 \delta   A_{\psi}^{(0)}   =  \frac{2 \pi  {\mathbbm Z}_6^{\psi}  }{6}  \;.    
\ee
The first term in (\ref{formula1}) is trivial, as
\be   \frac{1 }{ 8\pi^2}   \int   {\tr}  {\tilde F}^2  \in   {\mathbbm Z} \,, \qquad 
A_{\psi} =   d A_{\psi}^{(0)}\;, 
\ee
(which is  the standard gauge anomaly, reducing   $ U(1)_{\psi} \longrightarrow {\mathbbm Z}_6^{\psi}$).

Due to the second term  in (\ref{formula1}),   $\delta   A_{\psi}^{(0)} $ gets now  multiplied by  ($N=6$ here) 
\be    -   \frac{ 6  N }{ 8\pi^2}   \int    \big(B_{\rm c}^{(2)}\big)^2   =   -    6  N   \Big(\frac{1}{3}\Big)^2  {\mathbbm Z}  =   - 6   \,   \frac{2}{3}  {\mathbbm Z}\ . 
\ee
We see the  reduction of the  global chiral ${\mathbbm Z}_6^{\psi}$  symmetry
\be  \delta   A_{\psi}^{(0)}   =    \frac{2\pi   \ell}{6}\ , \qquad \ell=1,2,\ldots, 6
\ee
to  its subgroup  ($\ell=3,6$), 
\be
{\mathbbm Z}_6^{\psi} \longrightarrow {\mathbbm Z}_2^{\psi} \ .
\label{reduced}
\ee
Thus it is not possible that the vacuum of this system is confining, has a mass gap, and with no condensates breaking the
 ${\mathbbm Z}_6^{\psi}$ symmetry.

   What are the implications  of  (\ref{reduced})  on the phase of the theory?   First of all,  it
  implies a threefold vacuum degeneracy, under the  assumption that  the system confines (with mass gap) and that 
no massless fermions are present in the infrared,    on which  $\frac{{\mathbbm Z}_6^{\psi}}{{\mathbbm Z}_2^{\psi}}$ can act.
A natural assumption is that there are some condensates which  ''explain"  such a reduction of the symmetry in the infrared.  However, physics depends on
which condensates form. 
The simplest assumption is that a
bi-fermion condensate
\be  \langle  \psi \psi  \rangle    \sim \Lambda^3  \ne 0 \label{thefact}
\ee
 forms.   
As    $\psi   \in  {\underline {20}}$
a scalar  bi-fermion composite may  be in one of the irreducible representations  of $SU(6)$,  appearing on the right hand side of the composition-decomposition
 \be  \yng(1,1,1) \otimes \yng(1,1,1) =  \yng(1,1,1,1,1,1)\oplus \yng(2,1,1,1,1)\oplus +\ldots\;.
 \ee
 The most attractive channel is  the first, ${\underline 1}$, it vanishes by the  Fermi statistics. 
   This leaves us with  the  second best possibility  that  $\psi \psi$ in the adjoint representation  acquires  a VEV
    (dynamical Higgs mechanism)  \cite{Raby,BKS,BK}.

Note that even though  such a condensate should necessarily be understood as  a gauge-dependent form of some gauge invariant  VEV, it
 {\it  unambiguously}  determines  the breaking of global discrete chiral symmetry as  (\ref{reduced}),
  where the broken symmetry  $\frac{{\mathbbm Z}^{\psi}_6}{ {\mathbbm Z}^{\psi}_2}$ acts on the degenerate vacua, permuting them.  
  The  reason for this is that as the global symmetry group ${\mathbbm Z}_6^{\psi}$  commutes with the color $SU(6)$ a gauge transformation cannot undo the nontrivial transformation of the condensate under ${\mathbbm Z}_6^{\psi}$.

 Of course,  four-fermion, gauge-invariant condensates such as  
\be     \langle  \psi \psi \psi \psi   \rangle    \ne 0 \;, \qquad {\rm or} \qquad     \langle  {\bar \psi } {\bar \psi}  \psi \psi   \rangle    \ne 0  \;,    \label{ginvariant}
\ee
could  also form, first of which also breaks ${\mathbbm Z}^{\psi}_6$ as  in (\ref{reduced}).\footnote{We however do not share the view expressed in \cite{Yamaguchi} that
the gauge non-invariance of the bi-fermion composite $\psi\psi$    means  $ \langle  \psi \psi  \rangle =0\;;$    $ \langle  \psi \psi \psi \psi   \rangle    \ne 0$.
   }

 The  bi-fermion $\psi\psi$  condensate being 
  in the adjoint representation, it is possible that physics  in the infrared  is  described by full dynamical Abelianization   \cite{BKS,BK}.
The low-energy theory could be an Abelian $U(1)^5$ theory.     In such a case,
although the infrared theory may look trivial, there is a remnant of the ${\mathbbm Z}_6$ symmetry of the UV theory.
 Domain walls  which connect the three vacua   would exist, and  nontrivial infrared  $3$D physics can appear there.

   $SU(6)$   models with  $N_{\rm f} \ge 2$  have been studied also.  Physics implications from the mixed anomaly turn out to depend quite nontrivially on the value of  $N_{\rm f}$\cite{BKL1}.  

\subsubsection{$SU(N)$   models \label{sec:special}}

We next  consider   $SU(N)$ ($N$ general, even) theory,    with  left-handed fermions $\psi$  in the self-adjoint, totally antisymmetric
representation.  It exhibits some
interesting features  of the generalized anomalies.   

The first coefficient of the beta function is
\be  b_0=   \frac{ 11 N -  2 N_{\rm f} T_R }{3}\ .
\ee
The twice Dynkin index is given by 
\be    2 \, T_R =  { {  N-2}    \choose {\frac{N-2}{2}}} \ .\ee
See Table~\ref{dynkins}   for   $2 T_R$ and $d(R)$  for some even values of $N$ .
\be \begin{tabular}{|c |c cccc| }
\hline
$N$ & $4$ & $6$ & $8$ & $10$ & $12$     \\
 \hline
$2 \, T_R$ & $2$  & $6$ & $20$ & $70$ & $252$   \\
$d(R)$  &  $6$ &  $20$ &  $70$ &  $252$ & $924$  \\  
   \hline
\end{tabular}   \ .
\label{dynkins}
\ee
 We limit  ourselves to  asymptotically free theories   ($N \le 10$).

The system has  an exact 
1-form symmetry: 
\be     
{\mathbbm Z}_{\frac N2}^{\rm c} \,:\quad e^{ i \oint A } \to      e^{\frac{  2\pi i  }{ N}  k }   \,   e^{ i \oint A }   \ ,  \qquad        k=2,4,\ldots  N \ ,
\ee
  as well as  a   global  discrete symmetry:
\be {\mathbbm Z}_{2 T_R}\,:\quad   \psi \to  e^{\frac{2\pi i}{2 T_R}j}\, \psi \ ,  \qquad        j=1,2,\ldots 2 T_R\,.
\ee
By introducing  a 1-form gauge fields $\big(B_{\rm c}^{(2)}, B_{\rm c}^{(1)}\big)$  
\be          \frac{N}{2}   B_{\rm c}^{(2)} =d B_{\rm c}^{(1)}\ 
\ee
(see Sec.~\ref{Simplecases}),   
one arrives at the conclusion that   
the phase of the partition function is transformed by 
\be    - \frac{2\pi k  }{2 T_R}     2 T_R   \frac{4}{N} {\mathbbm Z}   =       - {2 \pi  k}  \frac{4}{N} {\mathbbm Z}\ , \qquad k=1,2,\dots, 2T_R\ ,
\ee
under ${\mathbbm Z}_{2 T_R}$.   
In other words,  ${\mathbbm Z}_{2 T_R}$  become  in  general  anomalous.
 The consequence of this mixed anomaly  
 however  depends on $N$
 nontrivially:
\begin{description}
\item[(i)]
For $N=4$,   the mixed anomaly vanishes:
\be   \frac{4}{N}=1\ . 
\ee
 \item[(ii) ]  For $N=4 \ell$, $\ell \ge 2$, instead, 
\be        \frac{4}{N}=      \frac{1}{ \ell}\ , 
\ee
and the discrete  symmetry breaking takes the form,   
\be    {\mathbbm Z}_{2T_R}^{\psi} \longrightarrow  {\mathbbm Z}_{\frac{2T_R}{\ell}}^{\psi}  \;.
\ee   
For  $N=4\ell$,  we note that  $2 T_R$  is an integer multiple of $\ell $.

  \item[(iii)] Finally,
  for  $N=4\ell +2$,  
\be  2 T_R   \cdot  \frac{4}{N}=     2 T_R   \cdot  \frac{2}{2 \ell+1}\ , 
\ee
thus  the breaking of the discrete  symmetry is 
\be    {\mathbbm Z}_{2T_R}^{\psi} \longrightarrow   {\mathbbm Z}_{\frac{2  T_R}{2\ell+1}}^{\psi}  \;.
\ee
Just for a check,   for $N=4\ell+2$, $2\ell+1$ is a divisor of  $2 T_R$ (see Appendix of \cite{BKL1}).

\end{description}

The  systematics of different cases, (i) $\sim$  (iii),  above, can be nicely  understood  in terms of  
the properties of the fractional instantons (torons) of this model.     See \cite{BKL1}.

\subsection{Adjoint QCD   \label{aqcd}}

$SU(N)$  theories with $N_{\rm f}$ Weyl fermions  $\lambda$  in the adjoint representation,   are sometimes called 
''the adjoint QCD".  These systems have been extensively  studied by using different  methods: 
 by semi-classical analysis \cite{Unsal2007},  by direct lattice simulations \cite{Debbio}, and more recently, by studying  the  mixed anomalies \cite{GKKS,ShiYon,AnbPop1}.  See  also  \cite{Shifman}, and more recent work  \cite{Popp,WanWang}.

Here  the color ${\mathbbm Z}_N^{\rm c}$     center 1-form symmetry is exact, which can be fully gauged.
The system has    also a nonanomalous ordinary ($0$-form)   discrete chiral symmetry, 
\be    {\mathbbm Z}_{2 N_{\rm f} N}^{\lambda}: \quad   \lambda \to  e^{\frac{2\pi i  }{2 N_{\rm f} N} k} \lambda\ , \qquad  k=1,2,\ldots,  2 N_{\rm f} N\;,\label{0formdisc}
\ee
as is well known.
A set of gauge fields are introced:
\begin{itemize}
\item $A_{\lambda}$: \,\, $ {\mathbbm Z}_{2 N_{\rm f} N}^{\lambda}$ \,\, 1-form gauge field,
     for     (\ref{0formdisc});
\item $B^{(2)}_{\rm c}$:    \,\,$\mathbb{Z}_{N}^{\rm c}$ \,\, 2-form gauge field.
\end{itemize}

${\mathbbm Z}_{2 N_{\rm f} N}^{\lambda}$ is found to  induces a phase change  in the partition function, 
\be    \left(   \frac{2 N  N_{\rm f}   }{8\pi^2}      \int   {\tr}_{} {\tilde F}^2 -  \frac{2 N^2  N_{\rm f}   }{8\pi^2}     \int   (B_{\rm c}^{(2)} )^2 \right)    \delta A_{\lambda}^{(0)}   \ ,
\ee
\be   \delta A_{\lambda}^{(0)}  \in     \frac{2 \pi i}{ 2 N N_{\rm f}}  {\mathbbm Z}\;.
\ee
The first term conserves  $ {\mathbbm Z}_{2 N N_{\rm f}}^{\lambda}$;   the second term 
\be \Delta S ( \delta A_{\lambda}^{(0)} )    \in     \frac{2 \pi i}{N} {\mathbbm Z}\ ,
\ee   
breaks  the chiral discrete symmetry further  as 
\be    {\mathbbm Z}_{2 N N_{\rm f}}^{\lambda} \longrightarrow      {\mathbbm Z}_{2 N_{\rm f}}^{\lambda}  \label{adjqcd}
\ee
as found in   \cite{GKKS,ShiYon,AnbPop1}.

The case of $SU(2)$, $N_{\rm f}=2$, is of particular interest.  In this case,
the discrete chiral symmetry ${\mathbbm Z}_{8}^{\lambda}$ is broken by the 1-form  ${\mathbbm Z}_{2}^{\rm c}$ gauging 
 as 
\be
 {\mathbbm Z}_{8}^{\lambda}\longrightarrow   {\mathbbm Z}_{4}^{\lambda} \ .   \label{1formgauging}
\ee
The  invariance of the standard $SU(2)$ theory
\be  \lambda \to  e^{\pm \frac{ 2\pi i}{8}} \lambda\ ,  \label{mixedBis}
\ee
 becomes anomalous.

What is the implication of these results on the physics in the infrared?   
A familiar lore about the infrared dynamics of this system is (for instance, see \cite{Shifman}) that a condensate
\be  \langle \lambda^{\{I} \lambda^{J\}} \rangle \ne 0 \ , \qquad    SU(2)_{\rm f}  \longrightarrow SO(2)_{\rm f}\;   \label{familiar}  \ee
($I,J=1,2$ being the flavor $SU(2)_{\rm f}$ indices)  forms.   That would lead to four-fold degenerate vacua, and in each of them, two  NG bosons.  

In an interesting work \cite{AnbPop1}  Anber and Poppitz have proposed that the system instead may develop a four-fermion condensate,
\be    \langle  \lambda \lambda \lambda \lambda  \rangle \ne 0\ , \quad {\rm with } \quad  \langle \lambda \lambda \rangle  =0\;.  \label{AnbPop}
\ee
Such a condensate  breaks  $ {\mathbbm Z}_{8}^{\lambda} $ spontaneously to  ${\mathbbm Z}_{4}^{\lambda}$,  
leaving only doubly degenerate $SU(2)_{\rm f}$ symmetric vacua (and no NG bosons).
 Massless baryons of spin $\tfrac{1}{2}$ 
 \be    B \sim   \lambda \lambda \lambda \;\label{AnbPop2}
 \ee
 (which is necessarily a doublet of the  unbroken $SU(2)_{\rm f}$)  should appear in the infrared spectrum  to 
 saturate all the conventional 't Hooft and Witten anomaly matching conditions.   
 The action of the broken $  {\mathbbm Z}_{8}^{\lambda}/ {\mathbbm Z}_{4}^{\lambda}$ is 
 a permutation between the two degenerate vacua,
 \be     \langle  \lambda \lambda \lambda \lambda  \rangle \to -  \langle  \lambda \lambda \lambda \lambda  \rangle\ .      \label{AnbPopZ2}
 \ee
As usual   in an anomaly-matching discussion one can tell some dynamical scenario 
is consistent, but not that such a vacuum is necessarily realized.
It remains to  establish which between   the familiar  $SO_f(2)$ symmetric vacuum and  the proposed   $SU(2)_{\rm f}$ symmetric one, 
 is actually realized.

The adjoint QCD  with  general $N$  reduces to ${\cal N}=1$ supersymmetric Yang-Mills theory,  for  the spacial case of $N_{\rm f}=1$.
A  great number of results  on  nonperturbative  aspects are known  \cite{AfDiSe,NSV,AmatiKMV,DaviesHollo}  there.  
Note that in this case   (\ref{adjqcd})  leads to  an $N$ fold vacuum degeneracy,   in  agreement  with the well-known Witten index of pure ${\cal N}=1$ $SU(N)$ Yang-Mills theory.  

One may also start from the ${\cal N}=2$ supersymmetric $SU(2)$ Yang-Mills theory,   where many exact results for the infrared physics are  known \cite{SW1,SW2,Tachikawa}. It can be deformed to ${\cal N}=1$ 
theory   by an adjoint-scalar mass perturbation, which yields  a confining, chiral symmetry breaking  vacua.
  Exact calculation of gauge fermion condensates $\brc \lambda \lambda\ckt$  from this viewpoint  can be found in   \cite{FinnelPouliot,RiccoK}. 
The pure ${\cal N}=2$ theory could  also  be perturbed directly to ${\cal N}=0$ \cite{Cordova}.  In principle, such an approach can give hints about $N_{\rm f}=2$ adjoint QCD, even  though it is not a simple  task to identify  correctly  the vacuum which can be reached by such a deformation.

\subsection{QCD with  ''tensor quarks"
\label{two}}

We now move to theories with matter fermions in $N_{\rm f}$ pairs of 
\be     \psi, {\tilde \psi }  =  \yng(2) \oplus   {\bar {\yng(2)} }    \label{symmquarks}
\ee
or 
\be 
   \psi, {\tilde \psi }  =   \yng(1,1) \oplus   {\bar {\yng(1,1)} } \;. \label{antisymmquarks}
\ee
For reference, the standard QCD quarks  are  in    $  \yng(1) \oplus   {\bar {\yng(1)} }$.

The first beta-function coefficient is
\be  b_0= \frac{11 N -   2 N_{\rm f}  (N \pm 2)}{3} \ .
\ee
As the $k= \frac{N}{2}$ element of the  center ${\mathbbm Z}_N$ does not act, there is a 
\be  {\mathbbm Z}_2^{\rm c}  \subset {\mathbbm Z}_N^{\rm c}
\ee
center symmetry.\footnote{This model has been considered by Cohen \cite{Cohen}, in particular in relation with such a center symmetry, and concerning the possible existence of an order parameter for confinement.}
Also   there is a  discrete axial symmetry subgroup
\be {\mathbbm Z}_{2 N_{\rm f} (N \pm 2)}^{\psi}\,:\quad    \psi   \to    e^{\tfrac{2\pi i  }{ 2 N_{\rm f} (N \pm 2)} } \, \psi\ ,\quad   {\tilde \psi }   \to    e^{\tfrac{2\pi i  }{ 2 N_{\rm f} (N \pm 2)} } \,{\tilde \psi } \,,
\ee
respected  by instantons.  The $\pm$ refer respectively to two types of models,   Eq.~(\ref{symmquarks}) and Eq.~(\ref{antisymmquarks}).

Let us study for simplicity the  $N_{\rm f}=1$ theory:  the analysis is similar to the cases discussed in the preceding sections.   
The  anomaly is given by 
\be
     \left(  -   \frac{2(N \pm 2)}{8\pi^2}       {\tr} {\tilde F}^2    +   \frac{2N (N \pm 2)}{8\pi^2}      (B_{\rm c}^{(2)} )^2    \right)    \delta A_{\psi}^{(0)}  \label{secondterm}  \ee
where 
\be    
   \delta A_{\psi}^{(0)}  \in \frac{2\pi }{2(N \pm 2)}   {\mathbbm Z}_{2(N \pm 2)}  \;.
\ee
Now
\be   \frac{2(N \pm 2)}{8\pi^2}       \int   {\tr}_{} {\tilde F}^2 \in  2 (N \pm 2)    {\mathbbm Z}  \ ,  \ee
 means that the first term  of (\ref{secondterm})  is trivial.   By using 
\be     \frac{1}{8\pi^2}  \int     (B_{\rm c}^{(2)} )^2 =   \frac{1}{4} {\mathbbm Z}  \ ,
\ee
the second term gives an anomaly
\be   A  =  2\pi  \frac{N}{4}  \,   {\mathbbm Z}  \; .
\ee
We find therefore no anomaly  for $N= 4\ell$;     for  $N= 4\ell + 2$,    the (1-form) gauging breaks the discrete symmetry as 
\be
   {\mathbbm Z}_{2 (N \pm 2)}^{\psi} \longrightarrow    {\mathbbm Z}_{N \pm 2}^{\psi}\ .
\label{bsr}
\ee
The assumption of  the ''quark condensate"
\be  \langle  \psi  {\tilde \psi} \rangle  \ne 0  \;,   \label{itself}
\ee
is consistent with (\ref{bsr}).
The bi-fermion condensate (\ref{itself}) however  breaks the discrete symmetry as
\be     {\mathbbm Z}_{2 (N \pm 2)}^{\psi}\longrightarrow   {\mathbbm Z}_{2}^{\psi}\ ,  
\ee
i.e.,    stronger than   suggested by (\ref{bsr}).

\subsection{Chiral theories with $ \tfrac{N-4}{k}  $   $\psi^{\{ij\}}$'s and    $\tfrac{N+4}{k}$   ${\bar \chi_{[ij]} }$'s 
}
\label{chiral}

Our next  theoretical laboratory is  chiral $SU(N)$ gauge theories with matter fermions in a complex representation,  $ \tfrac{N-4}{k}  $   $\psi^{\{ij\}}$'s and    $\tfrac{N+4}{k}$   ${\bar  \chi_{[ij]} }$,   or
 \be   \frac{N-4}{k}  \,\,   \yng(2)   \oplus   \frac{N+4}{k}     \,\,    {\bar  {\yng(1,1)} }  \ .   \ee
 $k$ is a common divisor of $(N-4, N+4)$ and  $N\ge 5$.  
Asymptotic freedom requirement 
\be     11 N -   \frac{2}{k} (N^2-8) >0\ , 
\ee
is compatible with various possible choices for  $(N, k)$.   We studied two  simple models   in \cite{BKL1}:
\begin{description}
  \item[(i)] $(N, k) = (6,2)$: $SU(6)$ with
  \be   \yng(2) \oplus    5\,\, {\bar {\yng(1,1)}}\ ;\label{model1}
  \ee 
  \item[(ii)]   $(N, k) = (8,4)$: $SU(8)$ with  
  \be      \yng(2) \oplus  3\,\, {\bar {\yng(1,1)}}\ .\label{model2}
  \ee
\end{description}
Below  we review  the results of the  analysis of the first model,  $SU(6)$ theory with matter fields   in   \,\underline{$21$}\,$\oplus$\,$5$\,$ \otimes $\,\underline{${15}$}$^{*}$.
The implications of the mixed anomalies turn out to be quite subtle even for this simple model,  as will be seen.

Classically the symmetry group  is 
\be SU(5) \times  U(1)_{\psi}\times U(1)_{\chi}\ . \ee 
The anomalies:
\bea U(1)_{\psi} - [SU(6)]^2 &=&\frac{T_{\tiny \yng(2)}}{T_{\tiny \yng(1)}}=N+2=8 \ , \nonumber \\
 U(1)_{\chi} - [SU(6)]^2&=&\frac{5 T_{\tiny \bar{\yng(1,1)}}}{T_{\tiny \yng(1)}}=5(N-2)=20 \ ,
\eea
fixes  the charges of  the nonanomalous  $U(1)_{\psi\chi}\subset  U(1)_{\psi}\times U(1)_{\chi} $  symmetry:
\be  (Q_{\psi}, Q_{\chi}) = (5,- 2)\ .\ee
There are also unbroken discrete groups:
\be U(1)_{\psi} \longrightarrow {\mathbbm Z}_8^{\psi}\ , \qquad U(1)_{\chi}\longrightarrow {\mathbbm Z}_{20}^{\chi} \ .\ee
By studying the overlap of $ {\mathbbm{Z}}_8^{\psi} \times  {\mathbbm{Z}}_{20}^{\chi} $  and $U(1)_{\psi\chi}$ one arrives at 
\be  \frac{U(1)_{\psi\chi}\times {\mathbbm Z}_8^{\psi} \times {\mathbbm Z}_{20}^{\chi} }{{\mathbbm Z}_{40}} \sim  U(1)_{\psi\chi}\times {\mathbbm{Z}}_4\;
 \label{global} \ee 
(see (\ref{quotient}) and (\ref{Z40Bis}) below).   Taking furthermore 
the color center  and $SU_f(5)$ center into account,  the true anomaly-free  symmetry  group is:
\be 
 \frac{SU(5) \times  U(1)_{\psi}\times U(1)_{\chi}}{{\mathbbm{Z}}_6^{\rm c} \times{\mathbbm{Z}}_5^f} \longrightarrow \frac{SU(5) \times  U(1)_{\psi\chi}\times {\mathbbm{Z}}_4}{{\mathbbm{Z}}_6^{\rm c} \times{\mathbbm{Z}}_5^f} \ . \label{full}
 \ee

From the consideration of the standard 't Hooft anomaly analysis and by the impossibility of finding an appropriate  set of massless baryons \cite{BKL1},  one concludes that if the system confines 
the global symmetry (\ref{global}) must be broken spontaneously, at least partially.
Therefore the question is whether 
  the 1-form gauging of a center symmetry can tell us anything useful.

    First of all,
  one finds that both ${\mathbbm Z}_8^{\psi} $  and  ${\mathbbm Z}_{20}^{\chi}$
are broken by the 1-form ${\mathbbm Z}_2^{\rm c} $ gauging: 
\be      {\mathbbm Z}_8^{\psi} \longrightarrow    {\mathbbm Z}_4^{\psi}\ , \qquad  {\mathbbm Z}_{20}^{\chi} \longrightarrow    {\mathbbm Z}_{10}^{\chi} \ .\label{suggests}
\ee
\be  \delta A_{\psi}^{(0)} =  \frac{2\pi k}{8}\ , \quad k=2,4,\ldots, 8\ ,  \qquad   \delta A_{\chi}^{(0)} = - \frac{2\pi \ell}{20}\ , \quad \ell=2,4,\ldots, 20\ .
\ee
In order for the system to ''match" the reduction of the symmetry (\ref{suggests}) in the infrared,  some condensates 
are expected to form. 
 A more careful analysis is,  however, needed to find out  which bifemion condensates actually  occur in the infrared, in order to be consistent with  the 
 systematics of the  mixed-anomalies.

The division by  ${\mathbbm Z}_{40}$,   
\be  \psi \to e^{  5   i \alpha} \psi\ , \qquad \chi \to e^{- 2 i  \alpha} \chi\ , 
\ee
\be      \alpha=   \frac{2\pi  k}{40}\ , \qquad k=1,2,\ldots, 40\ ,
\ee
 in the global symmetry group, (\ref{global}),  is  relevant.   
The quotient
\be    {\mathbbm Z}_4 \sim  \frac{  {\mathbbm Z}_{20} \times  {\mathbbm Z}_8}{{\mathbbm Z}_{40}}   \label{quotient}
\ee 
also forms a subgroup acting as
\be  \psi \rightarrow e^{2\pi i \frac{2k}{8}}\psi=e^{2\pi i \frac{k}{4}}\psi \;, \qquad   
\chi \rightarrow e^{-2\pi i \frac{5k}{20}}\chi=e^{-2\pi i \frac{k}{4}}\chi\;,   \label{Z40Bis}
\ee
or  
\be  \delta A_{\psi}^{(0)} =  \frac{2\pi k }{4}\ , \qquad   \delta A_{\chi}^{(0)} = - \frac{2\pi k }{4}\ , \qquad  k=1,2,3, 4\,.  \label{fromZ4}
\ee
The subtlety  is that 
${\mathbbm Z}_{40}$ remains nonanomalous, even after 1-form gauging of   ${\mathbbm Z}_2^{\rm c} $:
\be   -   \left(  8  \cdot   \frac{2\pi k }{8} -  20\cdot  \frac{2\pi k }{20} \right)  \,  
\frac{1}{8 \pi^2}    \left[   \int   {\tr}_{}  {\tilde F}^2  -  6   \, (B_{\rm c}^{(2)} )^2   \right]   =0\;.
\ee 
At the same time,     ${\mathbbm Z}_{4}$  itself  is affected by the gauging of the  center ${\mathbbm Z}_2^{\rm c} $ symmetry.  
 We find \cite{BKL1} the generalized  anomaly:
\be   -  3\cdot  2\pi k \,  
\frac{1}{8 \pi^2}    \left[   \int   {\tr}_{}  {\tilde F}^2  -  6   \, (B_{\rm c}^{(2)} )^2   \right]   =2\pi k  \cdot \left(   {\mathbbm Z} +    3\cdot 6\cdot  \frac{\mathbbm Z}{4} \right) \;.
\ee
Thus  ${\mathbbm Z}_{4}$   is reduced to   ${\mathbbm Z}_{2}$    ($k=2,4$).

These are   the fates of the  discrete symmetries 
\be    {\mathbbm Z}_{20} \times  {\mathbbm Z}_8 \sim    {\mathbbm Z}_{40} \times  {\mathbbm Z}_4
\ee
under the gauged 1-form center symmetry  ${\mathbbm Z}_2^{\rm c} $.
What do they imply on possible condensates such as 
\be    \psi \chi\;,  \qquad   \psi \psi,   \qquad \chi \chi\; 
\ee
?
The MAC criterion might suggest formation of condensates in one or more of the  channels
\bea & & A:  \qquad  \psi \left(\raisebox{-2pt}{\yng(2)}\right) \, \psi \left(\raisebox{-2pt}{\yng(2)}\right)   \quad  {\rm forming}  \quad   \,  \raisebox{-6pt}{\yng(2,2)}\;;
\nonumber \\  & & B:  \qquad  \chi \left(\bar{\raisebox{-9pt}{\yng(1,1)}}\right) \, \chi  \left(\bar{\raisebox{-9pt}{\yng(1,1)}}\right)  \qquad \ \   {\rm forming}  \quad \bar{\raisebox{-12pt}{\yng(1,1,1,1)}}\;;
\nonumber \\  && C:  \qquad   \psi  \left(\raisebox{-2pt}{\yng(2)}\right)   \, \chi \left(\bar{\raisebox{-9pt}{\yng(1,1)}}\right)  \quad \  \ {\rm forming  ~ adjoint ~ representation}\,\, .\label{condensates}
 \eea
 The one-gluon exchange strengths corresponding to these scalar composites  are   proportional to   $ \frac{16}{6},  \frac{28}{6}, \frac{32}{6} $,  respectively.   
 Of these,    the last  is the most attractive, and thus one might be tempted   to assume that the only condensate in the system   is 
\be  \brc (\psi \chi)_{\rm adj} \ckt \ne 0\;.  \label{only}
\ee
However,  the mixed anomaly analysis as sketched here  shows   that at least two different types of condensates must form in the infrared.  For instance the 
${\mathbbm Z}_{4}  \to  {\mathbbm Z}_{2}$  breaking would not be accounted for if   (\ref{only}) were the only condensate.  
See  \cite{BKL1} for more details.    One is led to conclude  that   two  or all of the condensates (\ref{condensates}), are generated by the system.

\section{Generalized anomalies and phases of the generalized  BY and GG models  \label{BYGG} }

In this section we review the generalized anomaly in a large class of chiral gauge theories, BY (Bars-Yankielowicz) and GG (Georgi-Glashow) models.
The procedure for computing   the anomaly  against  gauging color-flavor locked 1-form ${\mathbbm Z}_N$ symmetry has been exhibited   in Sec.~\ref{review2},  
by using the simplest of this class of models,  $\psi\eta$ model.  The  master formula found there, however, can be  applied to any of the  BY and GG models. 
 
\subsection{Bars-Yankielowicz models  \label{BYmodels}  }

The BY models  are  
  $SU(N)$ gauge theories with Weyl fermions 
\beq
   \psi^{ij}\,, \quad    \eta_i^A\, \,, \quad    \xi^{i,a}  
\eeq
in
\be       \yng(2) \oplus   (N+4+p) \,{\bar   {{\yng(1)}}}\;\oplus   p \,{   {{\yng(1)}}}\; .
\ee
The indices run as
\beq
\footnotesize  i,j = 1,\ldots, N\;,\quad A =1,\ldots , N+4+p  \;,\quad a =1,\ldots , p  \;.
\eeq
The  $\psi \eta$ model corresponds to  $p=0$.
The number of the extra fundamental pairs  $p$ is limited by $\frac{9}{2}N-3$ before asymptotic freedom (AF) is lost. 
The classical symmetry group is
  \be  SU(N)_{\rmc} \times  U(1)_{\psi} \times  U(N+4+p)_{\eta} \times  U(p)_{\xi}\;.  \label{groups}
    \ee
Strong anomaly breaks the symmetry group (\ref{groups}) to
  \bea
&&p=0:   \quad  SU(N)_{\rmc}  \times   SU(N+4)_{\eta}   \times  U(1)_{\psi\eta}\;,  \nn \\
&&p=1:    \quad  SU(N)_{\rmc}  \times   SU(N+5)_{\eta}  \times   U(1)_{\psi\eta}\times  U(1)_{\psi\xi}\;,  
\nn \\
&&p>1:   \quad  SU(N)_{\rmc}  \times   SU(N+4+p)_{\eta}  \times  SU(p)_{\xi}  \times  U(1)_{\psi\eta}\times  U(1)_{\psi\xi}\;,  
 \label{grouplocal}
    \eea
    where   the anomaly-free combinations are:
    \be
U(1)_{\psi\eta} : \qquad  \psi\to \rme^{\im (N+4+p)\alpha}\psi\;, \quad  \eta \to \rme^{-\im (N+2)\alpha}\eta\;, \label{upe}
\ee  
 with $\alpha \in \mathbbm{R}$, and 
    \be
U(1)_{\psi\xi} : \qquad \psi\to \rme^{\im p \beta}\psi\;, \quad  \xi \to \rme^{-\im (N+2)\beta}\xi\;, \label{upx}
\ee  
with $\beta \in \mathbbm{R}$.
The choice of these    two unbroken $U(1)$'s is arbitrary.   For example another $U(1)_{\eta\xi} $ 
\be
U(1)_{\eta\xi} : \qquad  \eta \to \rme^{\im p \gamma}\eta \;, \quad  \xi \to \rme^{-\im (N+4+p)\gamma}\xi\;,
\label{upex}
\ee  
with $\gamma \in \mathbbm{R}$ may be chosen.

   Table~\ref{suv} summarizes the charges.
\begin{table}[h!t]
  \centering 
  \small{\begin{tabular}{|c|c|c|c|c|c|  }
\hline
\su      &  $SU(N)_{\rm c}  $    &  $ SU(N+4+p)$    &  $ SU(p)$     &   $ {U}(1)_{\psi\eta}   $  &   $ {U}(1)_{\psi\xi}   $  \\
\hline 
\sbu  $\psi$   &   $ { \yng(2)} $  &    $  \frac{N(N+1)}{2} \cdot (\cdot) $    & $   \frac{N(N+1)}{2} \cdot (\cdot)  $  & $N +4 +p$ & $p $  \\
  $ \eta$      &   $  (N+4+p)  \cdot   {\bar  {\yng(1)}}   $     & $N  \cdot  {\yng(1)}  $     & $N (N+4+p) \, \cdot  (\cdot)   $   &$-(N+2)$&$0$\\ 
$ \xi$      &   $  p \cdot   {  {\yng(1)}}   $     & $N p \, \cdot  (\cdot)   $     &  $ N  \cdot   { {\yng(1)}} $ &$0$&$-(N+2)$ \\
\hline   
\end{tabular}}
  \caption{\footnotesize The multiplicity, charges and the representations.   $(\cdot)$ is a singlet representation.
}\label{suv}
\end{table}

The standard  't Hooft anomaly matching study, based on  the (perturbative) symmetry group,   (\ref{grouplocal}),  led to  the observation that the anomaly triangles associated with  (\ref{grouplocal})  can be all matched in the infrared, assuming confinement,  
no condensate formation,   and  assuming  a simple sets of massless composite fermions - baryons - saturating all the anomaly triangles.  This is highly  non trivial,  as seen  in  the summary given in    Appendix~\ref{conf3}.   

At the same time, the anomaly-matching equations are consistent with dynamics Higgs phase also,  where certain bi-fermion condensates form,  
which break the color and part of the flavor symmetries dynamically,   leaving still some non-trivial unbroken chiral symmetry  in the infrared.  See the review  in Appendix~\ref{Higgs3}.

It is thus  important  to find out  whether or not new, mixed type of anomalies  are present in the theories, and whether  
 more stringent anomaly constraints arise,    capable of discriminating these two dynamical possibilities.

As already emphasized,  the  study of the mixed anomalies require clarifying the global structure of the symmetry group,  {\it  beyond}  their local properties,
(\ref{grouplocal}).  The result of a detailed analysis done  in \cite{BKL4}  is  that the true symmetry group of the BY  model is   
\be  
SU(N)_{\rm c}\times   \frac{  SU(N+4+p) \times  SU(p) \times {\cal H}  }{ \mathbb{Z}_{N}  \times \mathbb{Z}_{N+4+p} \times \mathbb{Z}_{p}} \;,
\label{scvvv}
\ee
where  
\be   { \cal H } =   U(1)_1 \times  U(1)_2   \times  ({\mathbbm Z}_2)_F \;,
\ee
when $p$ and  $N$ are both   even. That is, 
 it has two disconnected components.  $U(1)_1$ and  $U(1)_2$  being  any two out of 
$U(1)_{\psi\eta}$,  $U(1)_{\psi\xi}$, and   $U(1)_{\eta\xi}$.

 When  $p$ and/or $N$ is odd, instead,  
\be  {  \cal H  } =   U(1)_1 \times  U(1)_2\;:
\ee
it has only one connected component.   
In these cases   symmetry group is connected, and 
perturbative (algebra) aspects of the  't Hooft anomaly triangles exhaust the UV-IR  anomaly matching conditions.   See \cite{BKL2}. 
Thus the most interesting BY   models are those with $p$ and  $N$ both even, to which  we turn now.

The fact that   
\be      {\mathbbm Z}_N \subset  U(1)_{\psi\eta} \times  U(1)_{\psi\xi}  \times ({\mathbbm Z}_2)_F   \;\label{Z2si}
\ee
 for $N$, $p$ both even,  can be shown explicitly, by choosing 
\be {  \alpha}=  \frac{ 2\pi}{N}  \;,   \quad    {  \beta}=    - \frac{ 2\pi}{N}   \;.
\label{thatis} 
\ee
We couple the system to the appropriate background gauge fields, 
\begin{itemize}
\item $A_{\psi\eta}$: $ U(1)_{\psi\eta}$    1-form gauge field, 
\item $A_{\psi\xi}$: $ U(1)_{\psi\xi}$    1-form gauge field, 
\item $A_2 $:  $(\mathbb{Z}_{2})_F
$ 1-form gauge  field, 
\item  $   {\tilde a}$: $U(N)_{\rm c} $ 1-form gauge field, 
\item $B^{(2)}_\rmc$: $\mathbb{Z}_{N}$ 2-form gauge field.
\end{itemize} 
Under the 1-form gauge transformation the fields transform as
\begin{align}
 B^{(2)}_\rmc &\to B^{(2)}_\rmc+{\diff} \lambda_\rmc\;,\qquad \ \, B^{(1)}_\rmc \to B^{(1)}_\rmc+{N}\lambda_\rmc\;, \nn \\ 
     {\tilde a}& \to   {\tilde a} +  \lambda_{\rm c}\;,  \qquad \qquad \   {\tilde F}({\tilde a}) \to    {\tilde F}({\tilde a}) + d \lambda_{\rm c}\;, \label{continuous11} \nn \\
  A_{\psi\eta} &\to  A_{\psi\eta}  -      \lambda_{\rm c}\;,\nn \\  
   A_{\psi\xi}& \to  A_{\psi\xi}  +    \lambda_{\rm c}\;,\nn \\  
    A_2& \to  A_2+ \frac{N}{2}  \lambda_{\rm c} \;.
  \end{align}
 The fermion kinetic terms are:   
\bea
&&   \overline{\psi}\gamma^{\mu}\left(\partial +\calR_{\rmS}(\widetilde{a})  + \frac{N+4+p}{2}  A_{\psi\eta}  +  \frac{p}{2}  A_{\psi\xi}   +    A_2    \right)_{\mu}P_\rmL\psi  + \nonumber\\
&& \overline{\eta}\gamma^{\mu}\left(\partial + \calR_{\rmF^*}(\widetilde{a})  -   \frac{N+2}{2}   A_{\psi\eta}     -   A_2    \right)_{\mu}P_\rmL\eta+ \nn  \\
 && \overline{\xi}\gamma^{\mu}\left(\partial + \calR_{\rmF}(\widetilde{a})  -    \frac{N+2}{2}  A_{\psi\xi}  +  A_2 \right)_{\mu}P_\rmL\xi\;.
\label{naive1}
\eea 

Knowing the  $({\mathbbm Z}_2)_F$ charges   $+1$, $-1$, $+1$ for the fermions $\psi$ $\eta$ and $\xi$, respectively, and their representations under $SU(N)$,   the master formula   (\ref{masterF}) gives straightforwardly the result for the mixed anomaly:  the result is
\be          N^2  \,      {1\over 8 \pi^2}  \int_{\Sigma^4} (B^{(2)}_\rmc)^2    \,   \frac{1}{2}  \delta A_2^{(0)}  =    N^2 \times \frac{\mathbbm Z}{N^2} \,  ({\pm \pi})
=\pm    \pi   \times  {\mathbbm Z}\;:   \label{asin1} 
\ee
     a   $({\mathbbm Z}_{2})_F  - [{\mathbbm Z}_{N}]^2$ mixed  anomaly.

One finds no $({\mathbbm Z}_{2})_F$ anomaly in the  IR, if the system is in the symmetric vacuum of Appendix.~\ref{conf3}.  
Such an inconsistency would be avoided, if one  assumed  instead that the system is in the dynamical Higgs phase  (Appendix.~\ref{Higgs3}), as  the  color-flavor locked 1-form symmetry  
would be spontaneously broken.

\subsection{Georgi-Glashow models}

The GG models have matter fermions in
\beq
   \chi^{ij}\,, \quad    \eta_i^A\, \,, \quad    \xi^{i,a}  \;,
\eeq
i.e., in  
\be       \yng(1,1) \oplus   (N-4+p) \,{\bar   {{\yng(1)}}}\;\oplus   p \,{   {{\yng(1)}}}\;,
\ee
where 
\beq
\footnotesize  i,j = 1,\ldots, N\;,\quad A =1,\ldots , N-4+p  \;,\quad a =1,\ldots , p  \;.
\eeq

The simplest of the  GG models (with $p=0$)  -  call it the $\chi \eta$ model -   can be analyzed 
following the same steps taken in the case of the  
 $\psi\eta$ model,   in Sec.~\ref{review2}.
 The (true) symmetry group of the $\chi\eta$ model is 
  \be
SU(N)\times G_{\rm f}\;, \qquad  G_{\rm f}=\frac{SU(N-4) \times  U(1)_{\chi\eta}  \times (\mathbb{Z}_2)_F } { \mathbb{Z}_{N}\times \mathbb{Z}_{N-4}}\;.
\label{symmetryNevenchi}
\ee
for even $N$.  
 $U(1)_{\chi\eta}$   and   $(\mathbb{Z}_2)_F$    act as  
\be  U(1)_{\chi\eta}\;: \qquad   \psi  \to    e^{ i  \tfrac{N-4}{2} \beta} \psi\;, \qquad     \eta  \to    e^{-  i  \tfrac{N-2}{2} \beta} \eta\;;  \label{Uchieta}
\ee
\be (\mathbb{Z}_2)_F\;: \qquad   \chi,  \eta  \to  -\chi, -\eta\;.    \label{fparityBis} 
\ee
The division by ${\mathbb{Z}}_{N}$ in (\ref{symmetryNevenchi})  is  due to  the equivalence relation
\be
(\rme^{\im \beta}, (-1)^n)\sim (\rme^{\im (\beta-\frac {2\pi}{ N})}, (-1)^n\rme^{\im\frac {2\pi}{ N}\frac{N}{2}})\,, \label{equiv11}
\ee
meaning  that $U(1)_{\chi \eta}$ gauge field $A$  and $(\mathbb{Z}_2)_F$ gauge field $A_2^{(1)}$  have charge $-1$ and  $\frac{N}{ 2}$, respectively.

By introducing  the gauge fields
\begin{itemize}
\item $A_2$: $ ({\mathbbm Z}_{2})_F$   1-form gauge field, 
\item $A$: $U(1)= U(1)_{\chi\eta}$ 1-form gauge field, 
\item $B^{(2)}_\rmc$: $\mathbb{Z}_{N}$ 2-form gauge field, 
\end{itemize}
the analysis follows step by step that done in the $\psi\eta$ model. 
     The result of the calculation, by use of the master formula  (Sec.~\ref{review2}),   is that there is a mixed  $(\mathbbm Z_2)_F-[{\mathbbm Z}_N]^2$  anomaly  in the UV, 
     \be     \Delta  S^{(4)}_{\rm UV}   =     \pm i \pi   {\mathbbm Z}\;:    \label{anomchieta}
     \ee
the partition function changes sign under  (\ref{fparityBis}).

Of course,    the "massless baryons"    lead to no anomalies   in the infrared. 
 The conclusion is that in the confining phase   (see Appendix~\ref{conf4})      the mixed   $(\mathbbm Z_2)_F-[{\mathbbm Z}_N]^2$   anomaly 
 does not UV-IR   match. 
 In other words  such a  symmetric confining phase  cannot be  the correct vacuum of the system.   There is no difficulty for the dynamical Higgs phase.  
The fact that in this particular  (and only)   case of the simplest of the GG model ($p=0$), the $\chi\eta$ model,     the confining, no-condensate  phase  (Appendix~\ref{conf2})  and the dynamical Higgs phase (Appendix~\ref{Higgs2}) {\it  happen}  to have the same global symmetry of the massless sector  does not necessarily imply  that these phases are the same phase  (see a more detailed discussion on this  issue in \cite{BKL5}).

More general GG models with $p \ne 0$  have also  been analysed in detail,  in  \cite{BKL2}.  The analysis  is similar to that done  for the $\chi\eta$ model  and for the general $BY$ models reviewed above.   
The conclusion is that the confining, symmetric vacuum  (Appendix~\ref{conf4})  is not consistent with the implications of the generalized anomalies.   The Higgs phase 
(Appendix~\ref{Higgs4})  seems to be perfectly consistent with these new constraints.

\section{Strong anomaly and phases of chiral gauge theories  \label{sec:stronganomaly}} 

Very recently the present authors have realized \cite{BKL5} that  strong anomaly, which plays a prominent role in the solution of the so-called $U(1)_A$ problem in QCD, 
can also  be significant in the study of the phases of chiral gauge theories, such as those discussed so far in this review.  More concretely,  
a key observation is that the well-known low-energy strong-anomaly effective action for QCD, which reproduces the 
effects of the strong anomaly in the low-energy effective action, has  a nontrivial implication on the symmetry breaking pattern itself.  Note that there is a subtlety in this argument,  as the well-known QCD effective sigma model action already assumes the standard chiral symmetry breaking 
into vectorlike residual symmetries (see  (\ref{vector}) below).  

The  criterion we adopt  to study chiral gauge theories of unknown low-energy symmetries and phases,   is  that it should be possible to write a low-energy strong-anomaly
local effective  action  by using the low-energy degrees of freedom (NG bosons and/or   massless  composite fermions) present in the assumed phase.

The simple form of such a low-energy action  we will find  in the dynamical Higgs phase, in contrast to the impossibility of writing analogous  terms with massless baryons only (in the confining phase), provides another, independent, indication that the first type of  phase 
(dynamical Higgs phase, characterized by certain bi-fermion condensates) 
represents a more plausible IR phase of this class of models.

Before considering the chiral gauge theories we discussed in the previous sections from this new angle,    we first review quickly 
what is known about the $U(1)_A$ problem in  the standard QCD \cite{WittenU1,VenezianoU1}.

\subsection{$U(1)_A$ problem and the $\theta$ dependence in QCD  \label{QCDAnomaly}}

In QCD the $U(1)_A$ symmetry is broken by the strong anomaly, and also spontaneously broken by the quark condensate  

\be       \brc {\bar {\psi}_L}  \psi_R \ckt    \sim -  \Lambda^3  \ne 0\;,
\ee
which
breaks the nonanomalous chiral  symmetry to its vectorlike subgroup,
 \be    SU(N_{\rm f})_L \times   SU(N_{\rm f})_R  \to  SU(N_{\rm f})_V\;.   \label{vector}
\ee
At this point one expects, besides the  NG bosons of the $SU(N_{\rm f})_A$ symmetry (the pions),
 another NG boson relative to $U(1)_A$, ($\eta$ or $\eta^{\prime}$)\footnote{Here $\eta, \eta^{\prime}$ are the  $SU(2)$ or $SU(3)$  singlet  pseudoscalar mesons of the real world, as in the Particle Data Booklet.   The reader will not confuse them with the Weyl fermion  in the chiral  $\psi\eta$ or $\chi\eta$ models being studied here.},    which would get mass due to the strong anomaly.  The chiral Lagrangian allows us to understand how this works  qualitatively, and quantitatively in the large $N$ limit.

To reproduce  the effects of the strong anomaly, the authors of \cite{Rosenzweig}-\cite{Nath} add to the standard chiral Lagrangian 
\be  L_0=    \frac{F_{\pi}^2}{2}     \Tr   \,   \de_{\mu}  U   \de^{\mu}  U^{\dagger}     +    \Tr  M \, U  +{\rm h.c.} +   \ldots \;, \qquad  U \equiv  {\bar \psi}_R  \psi_L
\ee
($F_{\pi}$ is the usual pion-decay constant), a new term
\be       {\hat L}  =     \frac{i}{2}  q(x)  \, \log \det  U / U^{\dagger}    +\frac{N}{a_0 F_{\pi}^2}  q^2(x)  - \theta \, q(x)\;,\label{QCDanomeff} 
\ee   
where $q(x)$ is the topological density 
\be q(x)  =  \frac{g^2}{32\pi^2}  F_{\mu\nu}^a  {\tilde F}^{a, \mu\nu}\;,      \label{topodens}
\ee
and  $a_0$ is some constant of the order of unity.
The variation of $\hat L$ under the action of $U(1)_A$
\be
\psi_L \to e^{i \alpha} \psi_L\;, \quad   \psi_R \to e^{- i \alpha} \psi_R\;, \quad U \to  e^{2 i \alpha} \, U
\ee
reproduces the variation of the phase of the partition function,
\be     \Delta  S  =   2 N_{\rm f}  \alpha  \int  d^4x   \frac{g^2}{32\pi^2}  F_{\mu\nu}^a  {\tilde F}^{a, \mu\nu}\;,
\ee
due to the strong anomaly.

The expression (\ref{QCDanomeff}) can be manipulated as shown in  \cite{Rosenzweig}-\cite{Nath}: integrating away the topological charge $q(x)$ one obtains an equivalent expression
\be   {\hat L}  =    -  \frac{F_{\pi}^2  \, a_0 }{4 N}   ( \theta -   \frac{i}{2}  \log \det U/U^{\dagger})^2\;, \label{analog}
\ee
which is well defined as 
\be      \brc U \ckt  \propto {\mathbf 1}    \ne 0\;.
\ee
Expanding (\ref{analog}) around this VEV,
\be    U \propto  e^{  i   \tfrac{ \pi^a  t^a}{F_{\pi}}   + i    \tfrac{ \eta \, t^0}{F_{\pi}^{(0)}  }   }  =    {\mathbf 1} +    i   \frac{ \pi^a  t^a}{F_{\pi}}   + i    \frac{ \eta \,  t^0}{F_{\pi}^{(0)}  }  +\ldots \;,
\ee
one finds the mass term for the would-be  NG boson, $\eta$.

The idea here is to reverse the logics:  one can actually  argue that the presence of such an effective action needed for reproducing the strong anomaly  {\it   implies} a nonvanishing condensate,   $\brc U \ckt =   \brc \bar \psi_R \psi_L \ckt \ne 0$, and hence,  indirectly,  also  the spontaneous breaking of {\it  nonanomalous}   chiral symmetry,    (\ref{vector}), 
affecting the low-energy physics.

Even if in QCD  there is a powerful direct argument \cite{Vafa:1983tf} for such a  vectorlike symmetry-breaking pattern, it is   
interesting to note that the requirement of faithfully representing the $U(1)_A$ anomaly in the infrared  seems to imply the same conclusion. 

It is this kind of consideration that has led recently the present authors to apply an analogous argument to chiral gauge theories  \cite{BKL5},  by requiring the 
effective low-energy theory to be able to express the strong anomaly appropriately. The following (Sec.~\ref{strongpsieta}-Sec.~\ref{stronganomGG}) is the review of some of the results found.

\subsubsection{ ${\cal N}=1$ supersymmetric theories}

 Before proceeding to the discussion of chiral gauge theories,  let us make a  brief comment on    {${\cal N}=1$  supersymmetric models.   In the context of ${\cal N}=1$  supersymmetric gauge theories, the strong-anomaly effective action is derived  by using  the so-called 
Veneziano-Yankielowicz   (VY) and  Affleck-Dine-Seiberg (ADS)    superpotentials 
\cite{VY,TVY,AfDiSe}.  They correctly reproduce  in the infrared effective theory the effects of instantons and supersymmetric Ward-Takahashi identities,  and embody   the  anomaly of \cite{KonishiAnom,KonishiShizuya}.  This last one, known as Konishi anomaly, has direct  implications on the vacuum properties of the theory under consideration. It is a straightforward  consequence of the strong anomaly, via supersymmetry  \footnote{The Konishi anomaly can also be viewed as representing an anomalous supersymmetry transformation law for some composite fields  \cite{KonishiAnom,KonishiShizuya}.}.  
  The VY  and ADS superpotentials are indeed crucial in determining the infrared dynamics  and phases  of the ${\cal N}=1$  supersymmetric gauge theories. For a review, see for instance  \cite{Amati:1988ft}.

\subsection{$\psi \eta$  model  and strong-anomaly   \label{strongpsieta}}

Let us apply a similar idea, i.e.  of  writing an effective action which reproduces the strong anomaly of the UV theory in the low energy theory,    to one of the simplest chiral gauge theories, the $\psi\eta$ model (see Sec.~\ref{review2}).  
Let us remind ourselves briefly  of  the symmetries of the model. At the infinitesimal level the quantum symmetry group of $\psi\eta$ is 
\be
SU(N+4)_\eta \times U(1)_{\psi\eta}\;,
\ee
while any combinations of $U(1)_\psi\times U(1)_\eta$ different form $U(1)_{\psi\eta}$ is broken by a strong anomaly. The low-energy effective action  must capture this strong anomaly correctly.

In  Sec.~\ref{review2} it was  shown that the $\psi\eta$ model cannot confine maintaining the full global symmetry unbroken.  
The system instead can break the gauge symmetry dynamically  (as well as part of  the global symmetry), and  a color-flavor locking condensate forms:
\be
\brc \psi^{ij}\eta^A_j \ckt
=\begin{cases}   c_{\psi\eta}\, \delta^{iA} \;,     \qquad  A =1, \dots N\;,\\ 0\;,   \qquad A=N+1, \dots N+4\;.\end{cases}  \label{Psieta}
\ee
Unlike  $ {\bar \psi_R} \psi_L$   in QCD,  $  {\tilde \phi}=  \sum_{k,j}^N   \psi^{k j} \eta_{j}^k\;,$    is not gauge invariant. 
It is convenient to re-express this condensate in a gauge invariant form, i.e.
\be      \det   U  \;,     \qquad   U_{k \ell}    \equiv   \psi^{k j} \eta_{j}^{\ell} \;.
\ee
Such a gauge invariant condensate is fully equivalent to (\ref{Psieta}). It causes  the breaking
\be
SU(N+4)\times U(1)\rightarrow  SU(N)_{_{\rm cf}}  \times  SU(4) \times U(1)^{\prime}\;,     \label{consistent}
\ee
(Appendix~\ref{Higgs1}).     $U(1)^{\prime}$  is the unbroken combination of  $U(1)_{\psi\eta}$  and    $U(1)_D$,  
where $U(1)_D$ is  $U(1)\subset SU(N+4)$ generated by $T_D={\rm diag}(4 \cdot \mathbbm 1_{N \times N}, -N \cdot \mathbbm 1_{4\times 4} )$.

At this point it is useful to look into the NG boson sector of the theory,  which leads to an apparent puzzle. 
From the symmetry breaking one expects to find $8N$ nonabelian NG bosons relative to   $\frac{SU(N+4)}{SU(N)\times SU(4)\times U(1)_D}$, interpolated in a gauge invariant fashion by
\be
\phi^A=(\psi^{ij}\eta_j^a)^*(T^A)^a_b (\psi^{ik}\eta^b_k)\;.
\ee
Here $T_A$ are the $8N$ broken generators that connect   the $N$ dimensional subspace (where $SU(N)$ acts)   and the $4$ dimensional one (where $SU(4)$ acts).

The problem emerges when one considers $U(1)$ NG boson(s).  
There is certainly a  physical massless NG boson, living in 
\be \frac{U(1)_D\times U(1)_{\psi\eta}}{U(1)'}\;.  \label{living} 
\ee  
The gauge-invariant field that interpolates it can be taken as the (imaginary part of the) condensate $\det\;  U$ itself.   
However, with the condensate  (\ref{Psieta}) alone, there is no space for another possible NG boson, associated with the symmetry breaking of an anomalous $U(1)$ symmetry (any generic combination of   $U(1)_{\psi}$ and  $U(1)_{\eta}$ other than  $U(1)_{\psi\eta}$ is in fact  spontaneously broken by ${\brc \psi\eta\ckt}$).
This would-be NG boson would get a mass from the strong anomaly, but in any case it needs to be described by an interpolating field (which?).  Another related fact is  that there is actually a particular anomalous symmetry (a special combination of     $U(1)_{\psi}$ and  $U(1)_{\eta}$)
\be
U(1)_A:\ \begin{cases}
	\psi\rightarrow e^{i\alpha}\psi\;,\\
	\eta \rightarrow e^{-i\alpha}\eta\;,\label{eq:u1a}
\end{cases}
\ee
which is not spontaneously broken by the  $\psi \eta$ condensate.  How would  such a symmetry manifest itself  in the infrared?   
These are the first hints that the description in terms of the condensate $\psi\eta$  (or  $\det \;U$)   is not a complete one.

Another reason to look for other condensates is that it is  not possible to write an effective Lagrangian which realizes all the (nonanomalous) global symmetries, with the composite field $\det U$ alone. For further details see \cite{BKL5}. 

With these considerations in mind, let us construct  the correct form of the strong-anomaly effective action systematically. We start from the very beginning,  
\be   {\cal L}  = -  \frac{1}{4} F_{\mu\nu} F^{\mu\nu}  +      {\cal L}^{\rm fermions}   \;,
\ee
\be     {\cal L}^{\rm fermions}   
= -i \overline{\psi}{\bar {\sigma}}^{\mu}\left(\partial +\calR_{\rmS}(a)   \right)_{\mu}  \psi\;  
- i   \overline{\eta}{\bar {\sigma}}^{\mu} \left(\partial +  \calR_{\rmF^*}(a) \right))_{\mu}   \eta\;, \label{naivepsipeta}
\ee
where $a$ is the $SU(N)$ gauge field, and  the matrix representations appropriate for $\psi$ and $\eta$ fields are indicated with $\cal R_{\rmS}$ and $\cal R_{\rmF^*}$. 
We change the variables by 
\be   {\cal L}  = -  \frac{1}{4} F_{\mu\nu} F^{\mu\nu}  +      {\cal L}^{\rm fermions}   +  \Tr [(\psi\eta)^* U]    +{\rm {\rm h.c.}   } 
+   { B}\,(\psi\eta\eta)^* +{\rm{\rm h.c.}}\;,    \label{above} 
\ee
where  $U$  is the composite scalars of $N\times (N+4)$ color-flavor mixed matrix form,
\be      \Tr [(\psi\eta)^* U]   \equiv    (\psi^{ij}  \eta_j^m)^*  U^{im}  \;,
\ee
and $ { B}$  are the baryons   $ B \sim \psi \eta \eta$,  
\be     B^{mn}=    \psi^{ij}  \eta_i^m  \eta_j^n \;,\qquad  m,n=1,2, \ldots, N+4\;,     \label{baryons00}
\ee
antisymmetric in  $m  \leftrightarrow n$.   In writing down the lagrangian (\ref{above})   we have anticipated the fact that these baryon-like composite fields, present in the Higgs phase together with the composite scalars $~\psi\eta$  (see Appendix~\ref{Higgs1}),    are also needed to write  down the 
strong-anomaly effective action. This allows us to dodge the problem about the NG bosons (and to break $U(1)_A$), as we are going to explain.

 Integrating $\psi$ and $\eta$ out, one gets    
\be   {\cal L}^{\rm eff}  = -  \frac{1}{4} F_{\mu\nu} F^{\mu\nu}  + \Tr  ({\cal D}  U)^{\dagger}    { \cal D} U     -i  \overline{ B}   \, { \bar {\sigma}}^{\mu}   \partial_{\mu} { B}      -  V\;.
\ee
The potential  $V$ is assumed to be such that its minimum is of the form: 
\be  \brc  U^{im}  \ckt = \, c_{\psi \eta} \,\Lambda^3  \delta^{i m }\;,   \qquad \quad \     i, m=1,2,\dots  N\;,     \label{condU}  
\ee
and contains the strong anomaly term,  
\be   V  =   V^{(0)} +    {\hat L}_{\rm an}\;.
\ee
${\hat L}_{\rm an}$ of the form,
\be       {\hat L}_{\rm an}=   { \const}  \,   \left[  \log   \left(  \epsilon \,  { B}  { B}\, { \det }  U \right)    -  \log \left(  \epsilon \, { B}  { B}  \det  U \right)^{\dagger}  \right]^2\;  \label{strong} \ee
which is analogue of (\ref{analog}) in QCD. 
The argument of the logarithm 
\be \epsilon  \, { B}  { B} \,  \det  U  \equiv    \epsilon^{m_1,m_2,  \ldots, m_{N+4}} \epsilon^{i_1, i_2, \ldots, i_N}
{ B}_{m_{N+1}, m_{N+2}}  { B}_{m_{N+3}, m_{N+4}}       U_{i_1 m_1}  U_{i_2 m_2} \ldots  U_{i_N m_N}\;, \label{insteadof}
\ee
is invariant under the full (nonanomalous) symmetry,
\be  SU(N)_{\rm c}\times SU(N+4)\times   U(1)_{\psi\eta}\;
\ee
as it should be. Moreover it contains $N+2$ $\psi$'s and $N+4$   $\eta$'s, the correct numbers of the fermion zeromodes in the instanton background: it corresponds to  a  't Hooft's instanton n-point function, e.g.,  
\be    \brc  \psi\eta\eta(x_1) \psi\eta\eta(x_2) \psi\eta(x_3) \ldots    \psi\eta(x_{N+2}) \ckt\;.
\ee

This effective Lagrangian is well defined only if the argument of the logarithm takes a VEV. In particular it is natural to assume
\be      \brc  \epsilon^{(4)}   { B}  { B}   \ckt \ne 0\;, \qquad   \brc    \det  U  \ckt  \ne 0\;,   \label{condenseBB}
\ee
where 
\be    \epsilon^{(4)}   { B}  { B}   =    \epsilon_{\ell_1  \ell_2  \ell_3  \ell_4}   { B}^{\ell_1 \ell_2}   { B}^{\ell_1 \ell_2}\;, \qquad 
\ell_i =  N+1,\ldots, N+4\;.      \label{asabove} 
\ee
As  
\be    \brc    \det  U  \ckt     \propto  {\mathbf 1}_{N\times N}\;
\ee
takes up all flavors up to $N$   (the flavor $SU(N+4)$ symmetry can be used to orient the symmetry breaking this way),  
$  { B}  { B}  $ must be made of  the four remaining flavors,  as  in (\ref{asabove}). 
These baryons were not among those considered in the earlier studies \cite{ADS,BKL4}, but  assumed to be massless here,
and indicated as $B^{[A_2 B_2]}$  in Table~\ref{SimplestBis}. This is possible because these fermions do not have any perturbative anomaly with respect to the unbroken symmetry group, $SU(N)\times SU(4)\times U(1)$: 't Hooft anomaly matching considerations cannot tell if they are massive or not, either the two options are possible.

Now we see how the apparent puzzle about the NG bosons  hinted at  above  is solved. We can define the interpolating fields of the two NG boson by expanding the condensates, 
\bea    && \det  U  =  \brc \det  U\ckt +  \ldots   \propto      {\mathbf 1}  +      \frac{i}{F_{\pi}^{(0)}}  \,  \phi_0   +\ldots \;;      \nonumber \\
&& \epsilon^{(4)}   { B}  { B}   =   \brc  \epsilon^{(4)}   { B}  { B} \ckt +  \ldots  \propto
{\mathbf 1}   +     \frac{i}{F_{\pi}^{(1)}}  \,  \phi_1 +\ldots  \;, 
\eea
(here
$F_{\pi}^{(0)}$ and 
$F_{\pi}^{(1)}$  are some constants with dimension of mass). Clearly in general the physical NG boson and the anomalous would-be NG boson will be interpolated by {\it two}   linear combinations of $\phi_0$ and $\phi_1$. The effective Lagraingian allows  us to fix these linear combinations.

Indeed, as the effective Lagrngian  (\ref{strong})   is invariant under the nonanomalous symmetry group, and in particular $U(1)_{\psi\eta}$ and $U(1)_D$ do not act on $\epsilon\; BB\; \det U$, the NG boson which appears in the strong-anomaly effective action as the fluctuation of $\epsilon BB \; \det U$, 
\be      {\tilde \phi} \equiv   N_{\pi}   \left[ \frac{1}{F_{\pi}^{(0)}}   \,   \phi_0 +   \frac{1}{F_{\pi}^{(1)}}  \,  \phi_1 \right] \;, \qquad   N_{\pi} =  \frac  { F_{\pi}^{(0)} F_{\pi}^{(1)}}{
	\sqrt{\big(F_{\pi}^{(0)}\big)^2 +  \big(F_{\pi}^{(1)}\big)^2 }}  \;, 
\ee
 cannot be the massless physical one: it is the would-be NG boson relative to the anomalous symmetry. Indeed the effective action provides a mass term for this NG boson. 
 
The orthogonal combination   
\be      {\phi} \equiv   N_{\pi}   \left[ \frac{1}{F_{\pi}^{(1)}}   \phi_0   -    \frac{1}{F_{\pi}^{(0)}}  \phi_1 \right] \;,  \label{physical}  \ee 
i.e. the interpolating field of the physical NG boson living in the coset (\ref{living}), remains massless.

Before we have included in the low-energy description some massless baryon which are neither required, nor excluded by the 't Hooft anomaly matching. Now one can see their ultimate fate using the strong-anomaly effective Lagrangian. In particular (\ref{strong}) contains a 4-fermion coupling between these baryons, which, plugging the VEVs  (\ref{condenseBB}), provides a mass term for them.

The last remark is that the strong-anomaly effective action does not depend on the absolute value of the condensates, $BB$ and $\det U$, separately.   This simply means that these symmetry considerations alone cannot determine the mechanism of condensation. In particular, even if    $\brc \det U \ckt \neq 0$ is somehow expected, a VEV for $BB$ is more surprising, and probably due to residual dipole interactions between the baryons. However a more in-depth study on how these two flat directions are lifted by quantum effects is needed to precisely understand how these two condensate form.

\subsection{Strong anomaly:   the generalized BY models \label{BY} }  

As the solution given above on the  $\psi\eta$ model is notably  subtle, one might wonder whether a similar mechanism is at work in the
 generalized  Bars-Yankielowicz models,  
an $SU(N)$ gauge theory with Weyl fermions 
\beq
\psi^{ij}\,, \quad    \eta_i^A\, \,, \quad    \xi^{i,a}  
\eeq
in the direct-sum  representation
\be       \yng(2) \oplus   (N+4+p) \,{\bar   {{\yng(1)}}}\;\oplus   p \,{   {{\yng(1)}}}\; .
\ee

Also in this case (Ref.~\cite{BKL4}  and Appendix~\ref{conf2}) the conventional 't Hooft anomaly matching equations allow a  chirally symmetric confining vacuum, with massless baryons
\be    
{({ B}_{1})}^{[AB]}=    \psi^{ij}   \eta_i^{A}  \eta_j^{B}\;,
\qquad {({ B}_{2})}^{a}_{A}=    \bar{\psi}_{ij}  \bar{\eta}^{i}_{A}  \xi^{j,a} \;
\qquad {({ B}_{3})}_{\{ab\}}=    \psi^{ij}  \bxi_{i,a}  \bxi_{j,b}  \;,
\label{baryons10}
\ee
(the first is anti-symmetric in $A \leftrightarrow B$ and the third is  symmetric in $a \leftrightarrow b$), saturating all conventional 't Hooft anomaly triangles. 
The study of the generalized anomaly in Sec.~\ref{BYmodels} has  however shown  that such a vacuum is not consistent. 

A dynamical Higgs phase with condensates
\bea
&&  \brc  U^{iB}  \ckt= \brc  \psi^{ij}   \eta_i^B \ckt =\,   c_{\psi\eta} \,  \Lambda^3   \delta^{j B}\ne 0\;,   \qquad   j,B=1,\dots,  N\;,     \nonumber \\
&&   \brc  V^{a A}  \ckt=    \brc  \xi^{i,a}   \eta_i^A \ckt =\,   c_{\eta\xi} \,  \Lambda^3   \delta^{N+4+a,  A}\ne 0\;,    \nonumber \\ && \qquad \qquad \qquad \qquad \qquad \qquad \qquad \qquad  a = 1,\dots, p\;,  \quad  A=N+5,\dots, N+ 4+  p \;, \nonumber \\     \label{BYvevs}
&&   \eea
and  with symmetry breaking 
\bea
&& SU(N)_{\rmc}  \times   SU(N+4+p)_{\eta}  \times  SU(p)_{\xi}  \times  U(1)_{\psi\eta}\times  U(1)_{\psi\xi} \nn \\
&&  \xrightarrow{\brc  \xi   \eta \ckt , \brc  \psi \eta \ckt}      SU(N)_{{\rm cf}_{\eta}}  \times   SU(4)_{\eta}  \times  SU(p)_{\eta\xi}  \times  U(1)_{\psi \eta}^{\prime} \times  U(1)_{\psi \xi}^{\prime} \;, 
\label{symbres}
\eea
is fully consistent with the gauging of the center symmetry  (Ref. \cite{BKL4} and Sec.~\ref{BYmodels}). 

A strong-anomaly effective action for the BY  theories can be constructed in a way similar  to  the $\psi\eta$  model. 
Instead of  (\ref{insteadof}),  one has now
\bea   &&    \epsilon  \left({{ B}_1}  {{ B}_1}  \det  U   \det V  \right)  
\equiv    \epsilon^{m_1,m_2,  \ldots, m_{N+4+p}} \epsilon^{i_1, i_2, \ldots, i_N}  \epsilon^{k_1, k_2, \ldots, k_p}   \times   \nonumber \\
&\times  &    { B}_1^{[m_{N+1}, m_{N+2}]}  { B}_1^{[m_{N+3}, m_{N+4}]}       U^{i_1 m_1}  U^{i_2 m_2} \ldots  U^{i_N m_N}
V^{m_{N+5}    k_1}     \ldots  V^{m_{N+4+p}  k_p}  \;.    \nonumber \\
\label{insteadofBis}
\eea

The rest of the analysis can be completed by closely following  that of the $\psi\eta$ model discussed in Sec.~\ref{strongpsieta}.
We skip the details of  the analysis.  Let us note only  that the strong anomaly effective action with such a logarithm,  
is perfectly consistent with, and perhaps implies,  the condensates, (\ref{BYvevs}): i.e., that  the system  is in dynamical Higgs phase, (\ref{symbres}).  
It is, instead,   not possible to rewrite a strong-anomaly effective action  with logarithmic argument  (\ref{insteadofBis}),  
in terms of  massless composite fermions (\ref{baryons10}) alone.

\subsection{Strong anomaly and  the  $\chi\eta$  model   \label{stronganomchieta}}

It is an interesting exercise to apply the same reasoning about the strong anomaly  to the $\chi\eta$ model. 
We will find that there are good analogies with the $\psi\eta$ case studied above, but also quite significant differences.

In this model any combination of $U(1)_\chi$ and $U(1)_\eta$, except $U(1)_{\chi\eta}$ (see Table~\ref{SimplestAgain2}), is anomalous, therefore some term similar to (\ref{QCDanomeff})   (in QCD) should appear.

In the dynamical Higgs scenario for the $\chi\eta$ model,  there are two bi-fermon condensates, 
\be  \brc  \chi^{ij} \eta_j^m  \ckt  = c_{\chi\eta}\, \delta^{im} \, \Lambda^3 \;, \qquad  i,m =  1,2,\ldots, N-4\;,   \label{chietacond}
\ee
and 
\be  \brc  \chi \chi \ckt \ne 0\;.
\ee
This implements a two-step  breaking, 
\bea        SU(N) \times    SU(N-4) \times U(1)_{\chi\eta}  &\xrightarrow{ \brc  \chi \eta \ckt}&     SU(N-4)_{\rm cf} \times  SU(4)_{\rm c} \times  U(1)^{\prime}
\nonumber \\   &\xrightarrow{ \brc  \chi \chi \ckt}& 
SU(N-4)_{\rm cf} \times  U(1)^{\prime}\;.    \label{symmetrychieta}
\eea

As before, in order to construct a fully consistent effective action, one should keep the full nvariance of the original theory,  either spontaneously broken or not.
 To do so, it is convenient to re-express the condensates  (\ref{chietacond})  in a gauge invariant way.   The answer is  to write a single gauge-invariant condensate
\bea      U &=&    \epsilon_{i_1 i_2 \ldots i_N}    \epsilon_{m_1 m_2 \ldots m_{N-4}}    (\chi \eta)^{i_1 m_1}    (\chi \eta)^{i_2 m_2} \ldots 
(\chi \eta)^{i_{N-4} m_{N-4}}    \chi^{i_{N-3}i_{N-2} }  \chi^{i_{N-1}i_{N} } \nonumber \\
&\sim&   \epsilon \, (\chi\eta)^{N-4} (\chi\chi)\;,   \label{naturalchieta} 
\eea
which encodes both of  the two (gauge depending) ones.

This choice suggests that   the correct  strong-anomaly  effective action  for the $\chi\eta$ model  is
\be         \frac{i}{2}    q(x)      \log    \epsilon  (\chi\eta)^{N-4} (\chi\chi)  + {\rm h.c.}\;,       \label{stronganomalychieta}    \ee    
where, again, $q(x)$ is the topological density defined in Eq. (\ref{topodens}). Clearly it  is by construction invariant under the whole (nonanomalous) symmetry group
\be 
SU(N)_{\rm c} \times  SU(N- 4) \times U(1)_{\chi\eta}\;.
\ee
Let us comment briefly some features that suggest that this is indeed the correct result.
\begin{itemize}
	\item The argument of the logarithmic function here matches the correct number of  the fermion zeromodes in the instanton background 
	($N_{\chi}= N-2$ and $N_{\eta}= N-4$);
	\item In contrast, there is no way  of writing  the strong anomaly effective action (\ref{stronganomalychieta})  in terms of the "baryons",   $B \sim \chi \eta \eta$, of the assumed confining, chirally symmetric phase (Appendix~\ref{conf2}).  No combination of the baryons can saturate the correct number of the fermion zeromodes,  {\it cfr}   (\ref{naturalchieta}).
\end{itemize}

This anomaly effective action  (\ref{stronganomalychieta}) agrees with  the one  proposed  by Veneziano \cite{Vene} for the case of  $SU(5)$,   and 
generalizes it  to all $SU(N)$ $\chi \eta$ models. A key observation we share with \cite{Vene} and generalizes to models with any $N$, is that this strong anomaly effective action,  which should be there  in the low-energy theory to reproduce correctly the (anomalous and nonanomalous) symmetries of the UV theory,  {\it  implies}   nonvanishing condensates, 
\be     \brc \chi\eta \ckt\ne 0\;, \qquad  \brc \chi\chi \ckt  \ne  0\;,     
\ee         
i.e.,  that the system is in dynamical Higgs phase, Appendix~\ref{Higgs2}.

Up to now the story has been very similar to the one about the $\psi\eta$ model. However there are some differences.
Differently from the $\psi\eta$ model, where the baryon condensate mush enter in the strong-anomaly effective action, here the structure of the effective action simplifies, and no baryon is needed.
Moreover, contrary to the $\psi\eta$ model,  the $\chi\eta$  system has no physical $U(1)$ NG boson: it is eaten by a color $SU(N)$ gauge boson.    However  
the counting of the broken and unbroken $U(1)$ symmetries  is basically similar in the two models.  Of the two nonanomalous 
symmetries ($U(1)_{\rm c}$ and $U(1)_{\chi\eta}$), a combination remains a manifest physical symmetry, and the other becomes the longitudinal part of a color   gauge boson.  Still another, anomalous,  $U(1)$ symmetry exists, which is any combination of $U(1)_{\chi}$ and  $U(1)_{\eta}$ other than  
$U(1)_{\chi\eta}$.  This symmetry is also spontaneously broken,  hence it  must be associated with a NG boson, even  though it will get mass by the strong anomaly.

As in the $\psi\eta$ model, one can describe this situation  explicitly, by expanding  the composite  $ \chi\eta$ and $\chi\chi$  fields around their VEV's,   
\bea    && (\det U)^{\prime} =  \brc   (\det U)^{\prime}  \ckt +  \ldots   \propto      {\mathbf 1}  +  i   \frac{1}{ F_{\pi}^{(0)}}  \, \phi_0^{\prime}    +\ldots \;;      \nonumber \\
&& \chi \chi   =   \brc  \chi \chi  \ckt +  \ldots  \propto 
{\mathbf 1}   + i  \frac{1}{ F_{\pi}^{(1)}}   \,   \phi_1^{\prime}   +\ldots  \;, 
\eea
where $ (\det U)^{\prime} $  is defined in the $N-4$ dimensional color-flavor mixed space, and  
\be   \chi\chi  \equiv     \epsilon_{i_1,i_2,i_3,i_4} \chi^{i_1 i_2} \chi^{i_3 i_4}\;, \qquad    N-3  \le  i_j  \le  N\;. 
\ee
Now  one can see that  the strong-anomaly effective action  (\ref{stronganomalychieta})    gives mass to  
\be      {\tilde \phi}^\prime    \equiv   N_{\pi}   \left[ \frac{1}{ F_{\pi}^{(0)}}     \phi_0^{\prime}  +   \frac{1}{ F_{\pi}^{(1)}} \phi_1^{\prime}  \right] \;, \qquad   N_{\pi} =  \frac  { F_{\pi}^{(0)} F_{\pi}^{(1)}}{
	\sqrt{(F_{\pi}^{(0)})^2 +  (F_{\pi}^{(1)})^2 }}  \;, 
\ee
whereas an orthogonal combination 
\be      {\phi}^\prime    \equiv   N_{\pi}   \left[ \frac{1}{ F_{\pi}^{(1)}}     \phi_0^{\prime}  -    \frac{1}{ F_{\pi}^{(0)}} \phi_1^{\prime}  \right] \; \ee
remains massless. The latter corresponds to the  potential NG boson which is absorbed by the color  $T_{\rm c}$  gauge boson. 

\subsection{Generalized  GG  models and strong anomaly  \label{stronganomGG}}

Let us now turn to the generalized GG models \cite{BKL4}, i.e. $SU(N)$ gauge theories with Weyl fermions 
\beq
\chi^{[ij]}\,, \quad    \eta_i^A\, \,, \quad    \xi^{i,a}  
\eeq
in the direct-sum  representation, 
\be       \yng(1,1) \oplus   (N- 4+p) \,{\bar   {{\yng(1)}}}\;\oplus   p \,{   {{\yng(1)}}}\; .
\ee
It turns out that the simple structure of the strong-anomaly effective action (\ref{stronganomalychieta}), which does not need bi-baryon condensate, works out also in this case:  
\be         \frac{i}{2}    q(x)      \log  \epsilon  \chi\chi    \det \,  (\chi\eta) \det  (\xi\eta)   + {\rm h.c.}\;,       \label{stronganomalyGG}    \ee    
where a shorthand notation
\bea   &&  \epsilon  \chi\chi    \det \,  (\chi\eta) \det  (\xi\eta)  =   \nonumber \\
&=&     \epsilon_{i_1 i_2 \ldots i_N}  \,   \epsilon_{k_1 k_2 \ldots k_p} \,  \epsilon_{m_1 m_2 \ldots m_{N-4+p} }  \nonumber \\
&\times&
(\chi \eta)^{i_1 m_1}    (\chi \eta)^{i_2 m_2} \ldots  (\chi \eta)^{i_{N-4} m_{N-4}}     ( \chi^{i_{N-3}i_{N-2} }  \chi^{i_{N-1}i_{N} })  
(\xi \eta)^{m_{N-3} k_{1}}    \ldots   (\xi \eta)^{m_{N-4+p} k_{p}}\;\nonumber \\
\eea
has been used.
The strong anomaly action (\ref{stronganomalyGG}) {\it  requires} the condensates 
\bea
&& \brc  \chi^{ij}   \eta_i^A \ckt  =\const. \,  \Lambda^3   \delta^{j A}\ne 0\;,   \qquad   j=1,\dots,  N-4\;,   \quad  A=1, \dots , N -4 \;,  \nonumber \\
&&  \brc  \xi^{i,a}   \eta_i^B \ckt =\const. \,  \Lambda^3   \delta^{N-4+a, B}\ne 0\;,   \qquad  a=1,\dots,  p \;,   \quad B=N-3,\dots,N-4+  p \;,\nonumber \\
&&  \eea
{\it  and} 
\be 
\brc       \chi^{j_1, j_2}  \chi^{j_3, j_4}   \ckt      =  \const. \,  \epsilon^{j_1, j_2, \ldots, j_4} \Lambda^3  \ne 0\;, \qquad   j_1,\ldots, j_4 =  N-3, N-2, \ldots, N\;.
\ee
Hearteningly,     these are   exactly    the set of  condensates expected to occur  in the  dynamical  Higgs phase of the  GG models  (Ref. \cite{BKL4} and Appendix~\ref{Higgs4}).

 \subsection{Strong anomaly in the chiral gauge theories considered in Sec.~\ref{Simplecases} }  
 
Up to now we have analysed the implications of the  strong anomaly  in models  discussed  in Sec.~\ref{review2} and in Sec.~\ref{BYGG}. In these models, being able to discriminate  between different types of phases (confinement with unbroken global symmetry versus dynamical Higgs phase) was clearly very  important, as 
the conventional 't Hooft anomaly-matching algorithm could  not tell us  which type of the vacua were the correct ones.
We found indeed that  both the  recent generalized anomaly study (Sec.~\ref{review2}  and Sec.~\ref{BYGG})  and strong anomaly  consideration (Sec.~\ref{strongpsieta} - Sec.~\ref{stronganomGG})   seem to favor the dynamical Higgs phase,  in all these models.   

Of course, consideration of the strong-anomaly effective action is relevant also in other models.  For illustration we discuss  below  a few models studied in  Sec.~\ref{Simplecases}.

\subsubsection{$SU(6)$ model with a single  fermion in  a self-adjoint antisymmetric representation   \label{sec:Yama}}  

Consider an  $SU(6)$ model 
with  a {\it single} left-handed fermion in the representation,  
\be   {\underline{20}} \, = \, \yng(1,1,1)\,
\ee
which was studied in \cite{Yamaguchi,BKL1}  and  reviewed in Sec.~\ref{sec:Yamaguchi} above. The lessons learned from the the gauging of the 1-form  $ {\mathbbm Z}_3^{\rm c}$  symmetry have been that  the nonanomalous ${\mathbbm Z}_6^{\psi}\,$   symmetry must break spontaneously as 
\be
{\mathbbm Z}_6^{\psi} \longrightarrow {\mathbbm Z}_2^{\psi} \;,  
\label{reducedBis}
\ee
implying  a three-fold vacuum degeneracy\cite{Yamaguchi,BKL1}.  This could either  be  because  of a four-fermion condensate \cite{Yamaguchi}
\be  \langle  \psi \psi   \psi \psi   \rangle    \sim \Lambda^6  \ne 0 \;,    \qquad   \langle  \psi \psi  \rangle =  0 \;,     \label{thefour}
\ee 
or  due to a gauge-symmetry breaking bi-fermion  condensate \cite{BKL1}
\be  \langle  \psi \psi  \rangle    \sim \Lambda^3  \ne 0 \;,    \label{thefactBis}
\ee
with $\psi\psi$  in the adjoint representation of $SU(6)$. Both scenarios are consistent.

Let us see if considerations on the strong-anomaly can clarify which scenario is actually realized. A particularly simple representation of the strong-anomaly is
\be         \frac{i}{2}    q(x)      \log      \psi \psi  \psi \psi  \psi \psi  + {\rm h.c.}\;.        \label{Yamaguchi}    \ee    
Based on our viewpoint  that the argument of the logarithmic function   acquires  a nonvanishing VEV,   the assumption of four-fermion condensate  (\ref{thefour}) appears to leads to a difficulty.     In constrast,   the assumption of bi-fermion condensates (\ref{thefactBis})  looks perfectly consistent, with 
\be    \langle   \psi \psi  \psi \psi  \psi \psi   \rangle  \sim      \langle  \psi \psi  \rangle^i_j  \langle  \psi \psi  \rangle^j_k  \langle  \psi \psi  \rangle^k_i  \ne 0\;.  
\ee

\subsubsection{Adjoint QCD with $N_{\rm c}=N_{\rm f}=2$ \label{adjointQCD}  }  

It is interesting to apply the same logics also to the adjoint QCD, previously analized from the point of view of generalized symmetries and their anomalies. In particular let us focus on the $N_{\rm c}=2$, $N_{\rm f}=2$ case, because of the renewed interest  in this particular model,  raised by the work of   Anber and Poppitz \cite{AnbPop1}.

The conventional thinking holds that a gauge invariant bi-fermion condensate
\be  \brc  \lambda \lambda \ckt  \ne 0   \label{Conv}  
\ee
forms, breaking the flavor symmetry as  $SU(2)_{\rm f} \to SO(2)_{\rm f}$, leading to $2$ NG bosons, and reducing the discrete ${\mathbbm Z}_8$ symmetry  to  ${\mathbbm Z}_2$ 
resulting four degenerate vacua.    The  Anber and Poppitz's proposal \cite{AnbPop1} is that,  instead,  the system might develop a four-fermion condensates but not bifermion condensate: 
\be     \brc  \lambda \lambda \lambda \lambda  \ckt  \ne 0\;, \qquad   \brc  \lambda \lambda \ckt = 0\;.    \label{AnbPo} 
\ee
 The discrete $\mathbbm Z_8$ symmetry is now broken to its $\mathbbm Z_4$ subgroup (therefore there are only two degenerate vacua). Massless baryons
\be    \sim  \lambda\lambda\lambda  
\ee
(necessarily a doublet of $SU(2)_{\rm f}$)   matches the UV-IR Witten anomaly of $SU(2)_f$.

As said above, the two possibilities are both consistent with the generalized 't Hooft anomaly matching,    therefore an indication from the strong-anomaly effective action would be very welcome. 

The analogue of (\ref{QCDanomeff}),   (\ref{stronganomalychieta})  and  (\ref{Yamaguchi}),  is   in this case, 
\be         \frac{i}{2}    q(x)      \log    \lambda \lambda  \ldots  \lambda  + {\rm h.c.}\;.          \label{hehe}
\ee
with eight $\lambda$'s  inside the argument of the logarithmic function.  Therefore, in contrast to what we saw in the preceding model  Sec.~\ref{sec:Yama},  our  strong-anomaly algorithm  does not seem to  be able to discriminate the two dynamical possibilities,    
(\ref{Conv}), or  (\ref{AnbPo}).   

Before concluding this section,  we note that   in the case with $N_{\rm f}=1$, arbitrary $N_{\rm c}$,   the adjoint QCD becomes ${\cal N}=1$ supersymmetric Yang-Mills theory.  The strong-anomaly effective action  (\ref{hehe}) 
with  $2N_{\rm f} N_{\rm c} =  2 N_{\rm c}$ ~$\lambda$'s,  reduces precisely to the Veneziano-Yankielowicz effective potential  \cite{VY}, implying  
$\brc  \lambda \lambda \ckt\ne 0$.    In this case  the assumption of the bi-fermion condensate, $\brc  \lambda \lambda \ckt\ne 0$, 
the breaking of the discrete symmetry  ${\mathbbm Z}_{2 N_{\rm c}} \to {\mathbbm Z}_2$, and the resulting 
$N_{\rm c}$ fold degeneracy of the vacua (Witten's index),  are generally accepted  as a well-established fact.

\section{Summary and discussion \label{conclude} }

We have reviewed in this article  the first applications of generalized anomalies   and discussed  what  the consequent stronger
anomaly matching conditions tell us about  various  chiral gauge theories based on $SU(N)$  gauge group. Our discussion was divided in two class of models.  In the first, the system has  a 1-form ${\mathbbm Z}_k$  symmetry ($k$ being a divisor of $N$) under which the matter fermions do not transform. The treatment in this case is relatively straightforward:  certain discrete symmetries, respected by instantons,  are often found to be further made anomalous, due to the fractional 't Hooft fluxes  accompanying the gauging of the discrete 1-form symmetries.  The discussion of Sec.~\ref{Simplecases} has illustrated that their consequences depend  nontrivially on the types of matter fermions present,  and  an interesting and rich variety of  predictions on the possible condensates,  symmetry breaking patterns  and phases, have been found. 

A second group of models (BY and GG  models) have a color-flavor locked ${\mathbbm Z}_N$  1-form symmetry, in which matter fermions transform together with the $SU(N)$ gauge field. A careful analysis of the global properties of the symmetry group is needed before actually introducing the gauging of this discrete 1-form symmetry.  This has been worked out in detail in all of the generalized  Bars-Yankielowicz and Georgi-Glashow models, and  the results of the analysis reviewed in Sec.~\ref{BYGG}.  A surprising implication is that, at least for even $N$ theories, color-flavor locked  ${\mathbbm Z}_2 - U(1) - {\mathbbm Z}_N$  
1-form symmetry does not allow its gauging  which is a new  kind of  't Hooft anomaly, and that this is consistent with the low-energy system being in dynamical Higgs phase characterized by certain bifermion condensates.

After the examination of the results from the generalized anomalies (involving the 1-form symmetries and gauging of some discrete center symmetries),  we changed the topics, and turned to the very recent observation \cite{BKL5} about the strong-anomaly associated effects.  This is the idea discussed in the context of the so-called $U(1)_A$ problem in QCD many years ago, but for some reason it was almost never applied to the discussion of the  physics of strongly-coupled chiral gauge theories.  
It is found that  the requirement that the massless degrees of freedom of  the hypothesized infrared phase  should be able to describe appropriately 
the strong anomaly   gives a rather clear indication on the physics of BY and GG models: the structure of the strong-anomaly effective action favors the dynamical Higgs vacua, against  the confining, fully flavor symmetric vacua, in agreement with 
the implications from the generalized anomaly matching algorithm, explored in the first part of this review.   

The fact that   both  mixed anomalies  \cite{BKL2,BKL4}  and  the strong-anomaly effective action \cite{BKL5}  imply dynamical Higgs phase in chiral BY and GG  models   is certainly not accidental.  Both arise by taking properly the strong chiral $U(1)$ anomalies into account.

These discussions, rather unexpectedly,  brought  us to note  certain analogies and contrasts between the strong-interaction dynamics of vector-like and chiral gauge theories \cite{BKL5}.
Let us  now compare the standard QCD  with $N_{\rm f}$  light flavors of  quarks and antiquarks,  and the 
$\psi\eta$, $\chi\eta$  models  as well as  more general   Bars-Yankielowicz and Georgi-Glasow models.

%
   In many senses,  the bifermion condensates such as  $U=  \psi\eta$  in the $\psi\eta$ model  (and $\chi\eta$,  $\chi\chi$ condensates in the $\chi\eta$  model), can be regarded as  a perfect analogue of the quark condensate
   $U= {\bar \psi}_R \psi_L  $ in QCD.   All of these composite scalars enter the strong-anomaly effective action in a similar way, as
   \be    {\hat L}  =  \frac{i}{2}    q(x)  \log   \det   U / U^{\dagger}\;, \qquad      q(x)  =  \frac{g^2}{32\pi^2}  F_{\mu\nu}^a  {\tilde F}^{a, \mu\nu}\;.
\label{anompsietaAll}
\ee 
(See Sec.~\ref{sec:stronganomaly} for more careful discussions.)
And in all cases  this implies condensation of  $\brc U \ckt \propto {\mathbf  1}$,  i.e.,  the color-flavor-locked Higgs phase in the $\psi\eta$ or $\chi\eta$ models on the one hand,   and 
the chiral-symmetry broken vacuum  in QCD., on the other.     

Another fact  pointing to a similarity between massless QCD and BY and GG models is the following. 
In general Bars-Yankielowicz  models (with $p$ pairs of additional matter fermions), we saw that  there are two natural bifermion condensate channels:
 \bea   &   \psi \Big( \raisebox{-3pt}{\yng(2)}\Big) \,  \eta \Big(\bar{\raisebox{-3pt}{\yng(1)}}\Big)      \qquad  \  & {\rm forming}    \qquad      \raisebox{-3pt}{\yng(1)}\;,  \nonumber    \\
{\rm and}  &   \quad   \xi \Big({\raisebox{-3pt}{\yng(1)}}\Big) \, \eta \Big(\bar{\raisebox{-3pt}{\yng(1)}}\Big)    \qquad   \ & {\rm forming}  \qquad  (\cdot)  \;:
 \eea
 the gluon-exchange strengths in the two channels  
are, respectively, proportional to \\
   $  -      \frac{ (N+2)(N-1)}{N} $  and  
    $ -       \frac{N^2-1}{N} $\;.
     The  $\psi \eta$ channel  is slightly more attractive;   the strength is however  identical in the large $N$ limit. Note  also  that   $  \xi \eta $ has  the same quantum numbers as  $ {\bar \psi}_R \psi_L$ in QCD.
     Similarly for the comparison between the  condensates,     $\brc \chi \eta \ckt$ and   $\brc  \xi \eta  \ckt$ in the Georgi-Glashow models.
These considerations,  based on rather na\"ive  MAC \cite{Raby}  idea and thus are not rigorous,  nevertheless  give supports to the idea that  the  quark condensates in QCD   and  the bifermions condensates in the chiral gauge theories under study in this review,  are really on a very similar footing.  

Of course,  there are significant  differences, or contrast,   in the vectorlike and chiral gauge theories.  
The quark condensate $\brc {\bar \psi}_R \psi_L\ckt $  is a color singlet,  $SU(N_{\rm f})_L\times SU(N_{\rm f})_R$ flavor matrix.
$\brc \psi\eta \ckt$  is  instead  in a color-flavor bifundamental form,   which means that it breaks the color completely, and reduces  (partially or totally) the unbroken flavor symmetry.
  The most important difference however is the existence of colored NG bosons in the $\psi\eta$  (or in the $\chi\eta$)  models.   It  means that these are coupled linearly to the gauge boson fields,    making them massive.   These processes are absent in QCD,  as  all NG bosons are color singlets. It is in this sense that  one talks about confinement phase in QCD,  in spite of the fact that the inter-quark confining strings can be broken  by the spontaneous 
quark-pair production from the vacuum.  

The mass spectra are also qualitatively different  in QCD and in the chiral gauge theories  discussed here. One is the presence of  certain degenerate massive vector bosons    (the color-flavor locked $SU(N)_{\rm cf} $ symmetry) found in the chiral gauge theories in the Higgs phase. But especially   the {\it  massless spectrum} exhibits striking differences.   In all chiral gauge theories studied  here,   it 
 contains in general {\it both} a number of composite fermions (baryons) as well as some composite scalars (pions),
  a feature  certainly not shared by massless QCD.   In other words, the way the chiral symmetries of the theory are realized in the IR, is notably different, in  vector-like and chiral gauge theories.

It is possible to see a closer analogy -  from a formal point of view - between the vector-like theories and chiral theories,  if one considers 
color superconductivity phase in the high-density region of QCD \cite{Alford1997,Alford1998}. The dynamics of QCD in that phase is believed to be such that some  colored di-quark condensates
    \be     \brc \psi_L  \psi_L  \ckt   \ne 0\;,   \qquad     \brc \psi_R  \psi_R  \ckt \ne 0\;.
    \ee
    form.
   In particular, in the case with $N_{\rm f}=3$ flavors  these are condensates of color-flavor diagonal form,  showing some similarity  to $\brc \psi\eta \ckt$  or  $\brc \chi\eta \ckt$ 
in the chiral theories discussed here. Of course,  the details of the dynamics will be quite different.

Summarizing,   the implications of new, mixed anomalies and the associated stricter anomaly-matching constraints reviewed in this article, and 
the consideration of the strong-anomaly effective actions,  together,  seem to allow us to get a clearer picture of the infrared dynamics of many strongly-coupled chiral gauge theories than before. It is to be seen whether some of these developments will turn out to be useful in a future effort to construct a realistic theory of Nature  beyond the standard Glashow-Weinberg-Salam-QCD model of the fundamental interactions.

\section*{Acknowledgments}  
 
  The work is supported by the INFN special 
research initiative grant, "GAST" (Gauge and String Theories).

\appendix

   \section{Chirally symmetric confining  $\psi\eta$ model
\label{conf1} } 
    
 An interesting possibility  pointed out for this model   is   that no condensates form, the system confines and the flavor symmetry remains  unbroken
     \cite{Dimopoulos:1980hn}.
    The candidate massless degrees of freedom in the IR  are
      \be     B^{[AB]}=    \psi^{ij}  \eta_i^A  \eta_j^B \;,\qquad  A,B=1,2, \ldots, N+4\;,     \label{baryons}
\ee
(baryons), 
antisymmetric in  $A \leftrightarrow B$.
All of  the $SU(N+4)_{\rm f}\times U(1)$ anomaly triangles  are saturated by  $ B^{[AB]}$, see  Table~\ref{Simplest0}. 
\begin{table}[h!t]
  \centering 
  \begin{tabular}{|c|c|c |c|c|  }
\hline
$ \phantom{{{   {  {\yng(1)}}}}}\!  \! \! \! \! \!\!\!$   & fields  &  $SU(N)_{\rm c}  $    &  $ SU(N+4)$     &   $ { U_{\psi\eta}}(1)   $  \\
 \hline 
  \phantom{\huge i}$ \! \!\!\!\!$  {\rm UV}&  $\psi$   &   $ { \yng(2)} $  &    $  \frac{N(N+1)}{2} \cdot (\cdot) $    & $   N+4$    \\
 & $ \eta^{A}$      &   $  (N+4)  \cdot   {\bar  {\yng(1)}}   $     & $N \, \cdot  \, {\yng(1)}  $     &   $  - (N+2) $ \\
   \hline     
 $ \phantom{ {\bar{   { {\yng(1,1)}}}}}\!  \! \! \! \! \!\!\!$  {\rm IR}&    $ B^{[AB]}$      &  $  \frac{(N+4)(N+3)}{2} \cdot ( \cdot )    $         &  $ {\yng(1,1)}$        &    $ -N    $   \\
\hline
\end{tabular}
  \caption{\footnotesize  Chirally symmetric ``confining"  phase of the  $\psi \eta $  model.   As in other Tables of the text, 
the multiplicity, charges and the representation are shown for each  set of fermions. $(\cdot)$ stands for a singlet representation.
}\label{Simplest0}
\end{table}

\section{Higgs phase of the $\psi\eta$ model  \label{Higgs1}  }

Another possibility for  the $\psi\eta$ model is  that a color-flavor locked phase appears \cite{ADS,BKS}, with 
\be    \brc  \psi^{\{ij\}}   \eta_i^B \ckt =\,   c_{\psi\eta} \,  \Lambda^3   \delta^{j B}\;,   \qquad   j, B=1,2,\dots  N\;,    \label{cflocking}
\ee
where  the symmetry is  reduced to 
\beq
SU(N)_{\rm cf} \times  SU(4)_{\rm f}  \times U(1)^{\prime} \,.   \label{isnotNew}
    \eeq
    A subset of the same baryons   ($B^{[A_1 B_1]}$ and   $B^{[A_1 B_2]}$ in the notation of Table~\ref{SimplestBis})
    saturate all of the triangles for  (\ref{isnotNew}), see Table~\ref{SimplestBis}.  
The  massless  degrees of freedom are  $\tfrac{N^2+7N}{2}$ massless baryons  $B^{[A_1 B]}$   and   $8N+1$  NG bosons.

   To reproduce correctly the strong anomaly in the IR, however,  another condensate $  \brc { B}  { B}   \ckt \sim  \brc  \psi \eta \eta \psi \eta \eta    \ckt $ 
 and another set of  massless baryon  $B^{[A_2 B_2]}$ (see Table~\ref{SimplestBis}) are needed.
This does not alter  neither the symmetry breaking pattern (\ref{isnotNew}), nor the anomaly matching. See Sec.~\ref{strongpsieta} for more detail.    
    \begin{table}[h!t]
  \centering 
  \begin{tabular}{|c|c|c |c|c|  }
\hline
$ \phantom{{{   {  {\yng(1)}}}}}\!  \! \! \! \! \!\!\!$   & fields   &  $SU(N)_{\rm cf}  $    &  $ SU(4)_{\rm f}$     &   $  U^{\prime}(1)   $  \\
 \hline
   \phantom{\huge i}$ \! \!\!\!\!$  {\rm UV}&  $\psi$   &   $ { \yng(2)} $  &    $  \frac{N(N+1)}{2} \cdot   (\cdot) $    & $   1  $    \\
 & $ \eta^{A_1}$      &   $  {\bar  {\yng(2)}} \oplus {\bar  {\yng(1,1)}}  $     & $N^2 \, \cdot  \, (\cdot )  $     &   $ - 1 $ \\
&  $ \eta^{A_2}$      &   $ 4  \cdot   {\bar  {\yng(1)}}   $     & $N \, \cdot  \, {\yng(1)}  $     &   $ - \frac{1}{2}  $ \\
   \hline 
   $ \phantom{{\bar{ \bar  {\bar  {\yng(1,1)}}}}}\!  \! \!\! \! \!  \!\!\!$  {\rm IR}&      $ B^{[A_1  B_1]}$      &  $ {\bar  {\yng(1,1)}}   $         &  $  \frac{N(N-1)}{2} \cdot  (\cdot) $        &    $   -1 $   \\
       &   $B^{[A_1 B_2]}$      &  $   4 \cdot {\bar  {\yng(1)}}   $         &  $N \, \cdot  \, {\yng(1)}  $        &    $ - \frac{1}{2}$   \\
       &   $B^{[A_2 B_2]}$      &  $   6  \cdot (\cdot )    $         &  $ {\yng(1,1)}  $        &    $0$   \\
\hline
\end{tabular}  
  \caption{\footnotesize   Color-flavor locked phase in the $\psi \eta$ model.
  $A_1$ or $B_1$  stand for the first $N$ flavors ($A_1,B_1=1,2,\ldots, N$),  whereas  $A_2$ or $B_2$ run  over 
   the rest of the flavor indices, $N+1,\ldots, N+4$.   Another set of potentially massless baryons   $B^{[A_2 B_2]}$ do not contribute to 
   $SU(N)_{\rm cf} \times  SU(4)_{\rm f} \times U^{\prime}(1)$ anomalies. 
   }\label{SimplestBis}
\end{table}

\section{Symmetric confining  phase for  the $\chi\eta$ model   \label{conf2}}

    Let us first examine the possible confining vacua, with full unbroken global symmetry \cite{Dimopoulos:1980hn}.
 The massless baryons needed for 't Hooft's anomaly matching  are   
 \be      B^{\{CD\}} = \chi_{[ij]} \,  {\eta}^{i\, C}    {\eta}^{j\, D}   \;, \qquad   C,D=1,2,\ldots (N-4)\;,  \label{massless}\ee
 symmetric in $C \leftrightarrow D$.   
\begin{table}[h!t]
  \centering 
  \begin{tabular}{|c|c|c |c|c|  }
\hline
 $ \phantom{{{   {  {\yng(1)}}}}}\!  \! \! \! \! \!\!\!$   & fields   &  $SU(N)_{\rm c}  $    &  $ SU(N-4)$     &   $ U(1)_{\chi\eta}   $  \\
 \hline 
  $ \phantom{{\bar{ \bar  {\bar  {\yng(1,1)}}}}}\!  \! \!\! \! \!  \!\!\!$  {\rm UV}&  $\chi$   &   ${\bar  { \yng(1,1)}}   $  &    $  \frac{N(N-1)}{2} \cdot (\cdot) $    & $N-4$    \\
& $ {\eta}^{A}$      &   $  (N-4)  \cdot   { {\yng(1)}}   $     & $N \, \cdot  \, {\yng (1)}  $     &   $ - (N-2)  $ \\
   \hline     
  \phantom{\huge i}$ \! \!\!\!\!$  {\rm IR}&    $ B^{\{AB\}}$      &  $  \frac{(N-4)(N-3)}{2} \cdot ( \cdot )    $         &  $ {\yng(2)}$        &    $ - N $   \\
\hline
\end{tabular}
  \caption{\footnotesize  Confinement and unbroken symmetry in the $\chi\eta$ model} 
  \label{01model}
\end{table}
%

\section{Higgs phase of the $\chi\eta$ model  \label{Higgs2} }

It was pointed out \cite{ADS,BKS} that this system may instead develop a color-flavor locked condensate,
\be   \brc   \chi_{[ij]} { \eta}^{B\, j}   \ckt  = \const   \, \Lambda^3 \delta_i^B \;, \qquad i, B=1,2,\ldots, N-4\;.  \label{cfl01}
\ee
The symmetry is broken to 
\beq
   SU(N-4)_{\rm cf}  \times  U(1)^{'}   \times   SU(4)_{\rm c}\,.
    \label{b270}
\eeq
The massless baryons (\ref{massless})  saturate all the anomalies associated with $SU(N-4)_{\rm cf}  \times  U(1)^{'}$.
There are  $\chi_{i_2 j_2}$  fermions which remain massless and strongly coupled to the $SU(4)_{\rm c}$.  We assume that 
$SU(4)_{\rm c}$ confines, and the condensate
\be  \brc  \chi \chi \ckt \ne 0\;, 
\ee
forms and $\chi_{i_2 j_2}$ acquire mass dynamically. 
Assume that the massless baryons are:
\be      B^{\{A B\}} = \chi_{[ij]} \,  {\eta}^{i\, A}   {\eta}^{j\, B}   \;, \qquad   A,B=1,2,\ldots (N-4)\;, 
\ee
 the saturation of all of the triangles associated can be seen in Table~\ref{SimplestAgain2}.  
\begin{table}[h!t]
  \centering 
  \begin{tabular}{|c|c|c |c|c|  }
\hline
$ \phantom{{{   {  {\yng(1)}}}}}\!  \! \! \! \! \!\!\!$   & fields     &  $ SU(N-4)_{\rm cf} $     &   $ U^{\prime}(1) $     &  $SU(4)_{\rm c}  $     \\
 \hline
   $ \phantom{{\bar{ \bar  {\bar  {\yng(1,1)}}}}}\!  \! \!\! \! \!  \!\!\!$  {\rm UV}&  $\chi_{i_1 j_1}$     &    $  {\bar  { \yng(1,1)}}   $    & $N$   &   $\frac{(N-4)(N-5)}{2}\cdot (\cdot)  $ \\
 &  $\chi_{i_1 j_2}$   &    $  4   \cdot {\bar  { \yng(1)}} $    & $\frac{N}{2}$   &   $ (N-4) \cdot {\bar  { \yng(1)}}   $     \\
 &$\chi_{i_2 j_2}$   &    $  \frac{4 \cdot 3}{2} \cdot (\cdot) $    & $0$    &   ${\bar  { \yng(1,1)}}   $     \\
& $ {\eta}^{i_1,A}$          & $\yng(2) \oplus \yng(1,1)$     &   $ - N $    &   $  (N-4)^2  \cdot  (\cdot)   $  \\
 & $ {\eta}^{i_2,A}$         & $4\, \cdot  \, {\yng (1)}  $     &   $ - \frac{N}{2} $     &   $  (N-4)  \cdot  \yng(1)  $  \\
   \hline 
     \phantom{\huge i}$ \! \!\!\!\!$  {\rm IR}&     $ B^{\{AB\}}$        &  $ {\yng(2)}$        &    $ - N $     &  $  \frac{(N-4)(N-3)}{2} \cdot ( \cdot )    $      \\
\hline
\end{tabular}
  \caption{\footnotesize  Color-flavor locking  in the $\chi\eta$ model.    The color index $i_1$ or $j_1$  runs up to $N-4$ and the rest is indicated by $i_2$ or $j_2$.}\label{SimplestAgain2}
\end{table}

 Complementarity \cite{Fradkin} appears to be working  here, in the sense that the massless sector of the dynamical  Higgs phase has the same 
 $SU(N-4)\times U(1)$ symmetry as in the  confining phase discussed in Appendix~\ref{conf2}.    See however a  discussion in \cite{BKL5}  on this point, which seems to point to the conclusion that the apparent complementarity (which occurs {\it  only}  in the $\chi\eta$ model,  but in none of other  BY and GG models)  is just a coincidence.

%
%

\section{Confining  symmetric phase of the BY  models     \label{conf3}}

%
%

   The candidate massless composite fermions  for the Bars-Yankielowicz  models are:
      \be    
 {({ B}_{1})}^{[AB]}=    \psi^{ij}   \eta_i^{A}  \eta_j^{B}\;,
\qquad {({ B}_{2})}^{a}_{A}=    \bar{\psi}_{ij}  \bar{\eta}^{i}_{A}  \xi^{j,a} \;,
\qquad {({ B}_{3})}_{\{ab\}}=    \psi^{ij}  \bxi_{i,a}  \bxi_{j,b}  \;,
\label{baryons10Bis}
\ee
the first is anti-symmetric in $A \leftrightarrow B$ and the third is  symmetric in $a \leftrightarrow b$, see  Table  \ref{sir}.
\begin{table}[h!t]
  \centering 
  \small{\begin{tabular}{|c|c|c |c|c| c|c| }
\hline
\su      &  $SU(N)_{\rm c}  $    &  $ SU(N+4+p)$    &  $ SU(p)$     &   $ {U}(1)_{\psi\eta}   $  &   $ {U}(1)_{\psi\xi}   $  \\
  \hline     
 \sbuu     $ {{ B}_{1}}$      &  $   \frac{(N+4+p)(N+3+p)}{2} \cdot ( \cdot )    $         &  $ {\yng(1,1)}$        &    $  \frac{(N+4+p)(N+3+p)}{2} \cdot ( \cdot )     $   &  $-N+p $ & $p$ \\
  \hline     
  \sbbu   $ {{ B}_{2}}$   &    $ (N+4+p) p\cdot ( \cdot )$     &       $  p\cdot   \bar{\yng(1)}$   &     $ (N+4+p)  \cdot {\yng(1)}$      & $-(p+2)$ & $-(N+p+2)$ \\
  \hline     
   \sbbu  ${{ B}_{3}}$     &  $ \frac{p (p+1)}{2}  \cdot ( \cdot )   $    &    $ \frac{p (p+1)}{2}  \cdot ( \cdot )   $       &    $ \bar{\yng(2)}$       & $N+4+p$ & $2N +4 + p$\\
\hline
\end{tabular}}
  \caption{\footnotesize  Chirally symmetric phase of the  BY    model. 
}\label{sir}
\end{table}
Explicit anomaly-matching checks can be found in e.g.,   \cite{BKL4}. 

\section{Higgs phase in the BY models   \label{Higgs3}}


Something nontrivial  happens in the dynamical Higgs phase for general BY models. i.e.,  with  $p > 0$.   There are now two  symmetry breaking channels, $\psi \eta$ and $  \xi \eta $. 
We assume that both  condensates occur as: 
\bea
&&  \brc  \psi^{ij}   \eta_i^B \ckt =\,   c_{\psi\eta} \,  \Lambda^3   \delta^{j B}\ne 0\;,   \qquad   j,B=1,\dots,  N\;,     \nonumber \\
&& \brc  \xi^{i,a}   \eta_i^A \ckt =\,   c_{\eta\xi} \,  \Lambda^3   \delta^{aA}\ne 0\;,   \qquad  a = 1,\dots, N\;,  \quad  A=N+1,\dots, N+ p \;, 
  \eea
where $\Lambda$ is the renormailization-invariant scale dynamically generated by the gauge interactions and $ c_{\eta\xi} ,  c_{\psi\eta} $ are coefficients both of order one. 
%
The resulting symmetry breaking pattern s
\bea
&& SU(N)_{\rmc}  \times   SU(N+4+p)_{\eta}  \times  SU(p)_{\xi}  \times  U(1)_{\psi\eta}\times  U(1)_{\psi\xi} \nn \\
&&  \xrightarrow{\brc  \xi   \eta \ckt , \brc  \psi \eta \ckt}      SU(N)_{{\rm cf}_{\eta}}  \times   SU(4)_{\eta}  \times  SU(p)_{\eta\xi}  \times  U(1)_{\psi \eta}^{\prime} \times  U(1)_{\psi \xi}^{\prime} \;.
\label{symbresBis}
\eea
The color gauge symmetry is completely (dynamically) broken, leaving  color-flavor diagonal  $SU(N)_{{\rm cf}_{\eta}} $ symmetry.
$U(1)_{\psi \eta}^{\prime}$ and $U(1)_{\psi \xi}^{\prime}$ are  combinations respectively  of $U(1)_{\psi \eta}$ (\ref{upe}) and $U(1)_{\xi \eta}$ (\ref{upx}) with  the element of $ SU(N+4+p)_{\eta}$ generated by 
\be   t_{SU(N+4+p)_{\eta}}= \left(\begin{array}{c|c|c}
(-\alpha (p+2) - p\beta) {\bf 1}_{N\times N}&&\\
\hline
&\frac{\alpha(N-p) - \beta p}{2}{\bf 1}_{4\times 4}&\\
\hline
&&(\alpha+\beta) (N+2) {\bf 1}_{p\times p}\\
\end{array}\right) \;.
\ee
Making the decomposition of the fields   one gets Table~\ref{brsuv}.
\begin{table}[h!t]
{  \centering 
  \small{\begin{tabular}{|c|c |c|c|c|c|  }
\hline
\su  &   $SU(N)_{{\rm cf}_{\eta}}   $    &  $SU(4)_{\eta}$     &  $ SU(p)_{\eta\xi}$ &   $  U(1)_{\psi \eta}^{\prime}$   &   $ U(1)_{\psi \xi}^{\prime}$ \\
 \hline
  \sbu $\psi$   &   $ { \yng(2)} $  &    $  \frac{N(N+1)}{2} \cdot   (\cdot) $    & $ \frac{N(N+1)}{2} \cdot   (\cdot)   $   &  $N +4 +p$ &  $p $  \\
   $ \eta_1$      &   $  {\bar  {\yng(2)}} \oplus {\bar  {\yng(1,1)}}  $     & $N^2  \cdot  (\cdot )  $     &   $ N^2  \cdot  (\cdot ) $    &$-(N +4 +p)$ & $-p $\\
    $ \eta_2$      &   $ 4  \cdot   {\bar  {\yng(1)}}   $     & $N  \cdot  {\yng(1)}  $     &   $ 4 N  \cdot  (\cdot ) $   & $-\frac{N+ p + 4}{2}$  & $-\frac{p}{2}$\\
  $ \eta_3$      &   $ p  \cdot   {\bar  {\yng(1)}}$     & $ N p  \cdot  (\cdot )  $     &   $N  \cdot  \bar{\yng(1)}  $    &$0$ &$N+2$ \\
    $ \xi$      &   $ p  \cdot   { {\yng(1)}}   $     & $N p  \cdot  (\cdot )  $     &   $ N  \cdot   { {\yng(1)}}   $   & $0$  & $-(N+2)$\\
\hline 
\end{tabular}}  
  \caption{\footnotesize   UV fieds in the BY  model, decomposed as a direct sum of the representations of the unbroken group of  Eq.~(\ref{symbresBis}). }
\label{brsuv}
}
\end{table}
\begin{table}[h!t]
{  \centering 
  \small{\begin{tabular}{|c|c |c|c|c|c|  }
\hline
 \su   &  $SU(N)_{{\rm cf}_{\eta}}   $    &  $SU(4)_{\eta}$     &  $ SU(p)_{\eta\xi}$ &   $ U(1)_{\psi \eta}^{\prime}  $   &   $ U(1)_{\psi \xi}^{\prime}$ \\
 \hline
      \sbbuu $ { B}_{1} $      &  $ {\bar  {\yng(1,1)}}   $         &  $  \frac{N(N-1)}{2} \cdot  (\cdot) $        &    $  \frac{N(N-1)}{2} \cdot  (\cdot) $     & $-(N +4 +p)$  &  $ -p $\\
   $ { B}_{2}  $      &  $   4 \cdot {\bar  {\yng(1)}}   $         &  $N  \cdot  {\yng(1)}  $        &    $4 N  \cdot  (\cdot ) $   & $-\frac{N+ p + 4}{2}$  & $-\frac{p}{2}$\\
\hline
\end{tabular}}  
  \caption{\footnotesize    IR fieds in the BY  model, the massless subset of the baryons in Tab.~\ref{sir} in the Higgs phase. }
\label{brsir}
}
\end{table}
The composite massless baryons are subset of  those in  (\ref{baryons10Bis}):
\bea
 & { B}_{1}^{[AB]} =  \psi^{ij}   \eta_{i}^{A}  \eta_{j}^{B} \;,  \qquad 
 { B}_{2}^{[AC]} =  \psi^{ij}   \eta_{i}^{A}  \eta_{j}^{\rm c}\;, & \nn \\
& A,B = 1, \dots, N \;,  \quad C=N+1, \dots, N+4 \;. &
\eea
%

\section{Confining symmetric phase  of the  GG    models     \label{conf4}}

The candidate massless composite fermions  for the generalized Georgi-Glashow models are:
  \be    
 {({ B}_{1})}^{\{AB\}}=    \chi^{ij}  \eta_i^{A}  \eta_j^{B} \;,
\qquad {({ B}_{2})}^{a}_{A}=    \bar{\chi}_{ij}  \bar{\eta}^{i}_{A}  \xi^{j,a} \;,
\qquad {({ B}_{3})}_{[ab]}=    \chi^{ij}  \bxi_{i,a}  \bxi_{j,b} \;,
\label{baryons20}
\ee
the first  symmetric in $A \leftrightarrow B$ and the third anti-symmetric in $a \leftrightarrow b$.
All anomaly triangles  are saturated  by these candidate massless composite fermions
 (Table~\ref{air} vs Tab.~\ref{sua}).  For explicit matching equations, see e.g., \cite{BKL4}. 
\begin{table}[h!t]
  \centering 
  \small{\begin{tabular}{|c|c|c |c|c| c|c| }
\hline
\su      &  $SU(N)_{\rm c}  $    &  $ SU(N-4+p)$    &  $ SU(p)$     &   $ {U}(1)_{\chi\eta}   $  &   $ {U}(1)_{\chi\xi}   $  \\
  \hline     
 \sbbu      $ { B}_{1}$   &  $  \frac{(N-4+p)(N-3+p)}{2} \cdot ( \cdot )    $         &  $ {\yng(2)}$        &    $  \frac{(N-4+p)(N-3+p)}{2} \cdot ( \cdot )    $   &  $-N+p $ & $p$ \\
  \hline     
 \sbbu    $ { B}_{2}$  &   $ (N-4+p) p \cdot ( \cdot )  $   &        $ p \cdot  \bar{\yng(1)}$   &     $  (N-4+p) \cdot  {\yng(1)}$     & $-(p-2)$ & $-(N+p-2)$\\
  \hline     
   \sbbuu  $ { B}_{3}$   &    $  \frac{p (p-1)}{2}  \cdot ( \cdot )  $  &      $ \frac{p (p-1)}{2}  \cdot ( \cdot )  $ &    $ \bar{\yng(1,1)}$   & $N-4+p$ & $2N -4 +p$\\
\hline
\end{tabular}}
  \caption{\footnotesize  IR massless fermions in the chirally symmetric phase of the  GG  model. 
}\label{air}
\end{table}

\section{Higgs phase in the GG   models     \label{Higgs4} }

 The Higgs phase of the GG model may be described by either of the  two possible bifermion  channels $\chi \eta$ and $  \xi \eta $.
We assume that
 both  occur 
\bea
&& \brc  \chi^{ij}   \eta_i^A \ckt  =\,   c_{\chi\eta} \,  \Lambda^3   \delta^{j A}\ne 0\;,   \qquad   j=1,\dots,  N-4\;,   \quad  A=1, \dots , N -4 \;,  \nonumber \\
&&  \brc  \xi^{i,a}   \eta_i^B \ckt =\,   c_{\eta\xi} \,  \Lambda^3   \delta^{aB}\ne 0\;,   \qquad  a=1,\dots,  p \;,   \quad B=N-4+1,\dots,N-4+  p \;. \nonumber \\
\eea
The  symmetry-breaking pattern is
\bea
&& SU(N)_{\rmc}  \times   SU(N-4+p)_{\eta}  \times  SU(p)_{\xi}  \times  U(1)_{\chi\eta}\times  U(1)_{\chi\xi} \nn \\
&&  \xrightarrow{\brc  \xi   \eta \ckt , \brc  \chi \eta \ckt}      SU(4)_{{\rm  c}}  \times  SU(N-4)_{{\rm cf}_{\eta}}  \times   SU(p)_{\eta\xi}  \times  U(1)_{\chi \eta}^{\prime} \times  U(1)_{\chi \xi}^{\prime} \;.
\label{symbrea}
\eea

The color gauge symmetry is partially (dynamically) broken, leaving  color-flavor diagonal  global $SU(N-4)_{{\rm cf}_{\eta}} $ symmetry and an $SU(4)_{{\rm  c}}$ gauge symmetry.
$U(1)_{\chi \eta}^{\prime}$ and $U(1)_{\chi \xi}^{\prime}$ are a combinations respectively  of $U(1)_{\chi \eta}$ (\ref{uce}) and $U(1)_{\chi \xi}$ (\ref{ucx}) with  the elements of $SU(N)_{\rmc} $ and $ SU(N-4+p)_{\eta}$ generated by:
\bea  
&& \qquad  t_{SU(N)_{\rm c}}=  \left(\begin{array}{c|c}
2 \frac{  \alpha  (N-4+p ) +\beta p }{N-4}  {\bf 1}_{(N-4)\times (N-4)}&\\
\hline
& - \frac{  \alpha  (N-4+p ) +\beta p }{2}  {\bf 1}_{4\times 4}\\
\end{array}\right) \ , \nn \\
&& t_{SU(N-4+p)_{\eta}}=  \left(\begin{array}{c|c}
-\frac{p (\alpha + \beta ) (N-2 )}{N-4} {\bf 1}_{(N-4)\times (N-4)}&\\
\hline
&  (\alpha + \beta ) (N-2 )  {\bf 1}_{p\times p}\\
\end{array}\right) \;.  
\eea
Making the decomposition of the fields in the direct-sum representations in  the subgroup one arrives at  Table~\ref{brauv}.
\begin{table}[h!t]
{  \centering 
  \small{\begin{tabular}{|c|c |c|c|c|c|  }
\hline
\su       &    $SU(N-4)_{{\rm cf}_{\eta}}   $  &  $SU(4)_{{\rm c}}$ &  $ SU(p)_{\eta\xi}$ &   $  U(1)_{\chi \eta}^{\prime}$   &   $ U(1)_{\chi \xi}^{\prime}$ \\
\hline
\sbbuu $\chi_1$     &       $ { \yng(1,1)} $ & $  \frac{(N-4)(N-5)}{2} \cdot   (\cdot) $  & $ \frac{(N-4)(N-5)}{2} \cdot   (\cdot)   $   &  $\frac{(N-4+p) N}{(N-4)}    $  &  $p \frac{N}{N-4}$  \\
$\chi_2$                &   $ 4 \cdot { \yng(1)} $  &    $  (N-4)  \cdot    { \yng(1)} $    & $ 4(N-4) \cdot   (\cdot)   $   &$\frac{(N-4+p) N}{2(N-4)}  $  &  $\frac{p N}{2(N-4)}$  \\
$\chi_3$                &    $  6  \cdot   (\cdot) $   &   $ { \yng(1,1)} $   & $ 6  \cdot   (\cdot)   $   &  $0$  & $0$  \\
$ \eta_1$                 &  $  {\bar  {\yng(2)}} \oplus {\bar  {\yng(1,1)}}  $     &   $(N-4)^2  \cdot   (\cdot )    $   &   $ (N-4)^2  \cdot  (\cdot ) $   & $ -\frac{(N-4+p) N}{(N-4)} $  & $-\frac{p N}{N-4} $\\   
$ \eta_2$             & $ p  \cdot \bar{\yng(1)} $      &   $p(N-4)  \cdot  (\cdot )  $      &   $ (N-4)  \cdot \bar{\yng(1)} $    &$-2 -2 \frac{p}{N-4}$ & $N-2 -\frac{2 p}{N-4}  $\\
$ \eta_3$              &   $  4   \cdot   {\bar  {\yng(1)}}$     & $ (N-4)  \cdot   {\bar  {\yng(1)}} $     &   $ 4  (N-4)     \cdot   (\cdot )  $    &$-\frac{(N-4+p) N}{2(N-4)} $ &$ -\frac{p N}{2(N-4)} $ \\
$ \eta_4$              &   $ 4 p \cdot  (\cdot ) $     & $ p \cdot   {\bar  {\yng(1)}}   $     &   $4  \cdot \bar{\yng(1)}   $    &$\frac{N-4+p }{2}$ &$N-2  + \frac{p }{2}$ \\
$ \xi_1$                &   $  p  \cdot   { {\yng(1)}}    $     & $p(N-4)   \cdot  (\cdot )  $     &   $ (N-4)   \cdot   { {\yng(1)}}   $   & $2 + 2\frac{p}{N-4}$  & $-(N-2)+\frac{2p}{N-4}$\\
$ \xi_2$                &   $4 p   \cdot  (\cdot )   $     & $  p  \cdot   { {\yng(1)}}    $     &   $ 4  \cdot   { {\yng(1)}}   $   & $-\frac{N-4+p}{2}$  & $-(N-2)-\frac{p}{2}$\\
\hline 
\end{tabular}}  
  \caption{\footnotesize   UV fieds in the GG   model, decomposed as a direct sum of the representations of the unbroken group of  Eq.~(\ref{symbrea}). }
\label{brauv}
}
\end{table}

\begin{table}[h!t]
{  \centering 
  \small{\begin{tabular}{|c|c |c|c|c|  }
\hline
\su     &    $SU(N-4)_{{\rm cf}_{\eta}}   $    &  $ SU(p)_{\eta\xi}$ &   $  U(1)_{\chi \eta}^{\prime}$   &   $ U(1)_{\chi \xi}^{\prime}$ \\
 \hline
\sbbu $ { B}   $          &  $  {\bar  {\yng(2)}}  $     &   $\frac{(N-4)(N-3)}{2}\  \cdot  (\cdot ) $   &  $ -\frac{(N-4+p) N}{(N-4)} $  & $-\frac{p N}{N-4} $\\
  \hline 
\end{tabular}}  
  \caption{\footnotesize   IR fied in the GG  model in the dynamical Higgs phase. }
\label{brair}
}
\end{table}
The composite massless baryons are subset of those in (\ref{baryons20}): 
\bea
 && { B}^{\{AB\}} =  \chi^{ij}   \eta_{i}^{A}  \eta_{j}^{B} \;,  \qquad  A,B = 1, \dots, N-4 \;.   \label{baryonbra}
\eea
In the IR these fermions  saturate all the anomalies of the unbroken chiral symmetry. 

There is a novel feature in the GG   models, which is not shared by the BY   models.   
There is an  unbroken strong gauge symmetry $SU(4)_{{\rm c}}$, with a  set of  fermions, 
\be  \chi_3\;,\quad \chi_2\;,\quad  \eta_3\;,\quad \eta_4\;,\quad \xi_2\;,\label{these}
\ee
charged with respect to it.    The pairs  $\{ \chi_2\;,\,  \eta_3 \}$  and   $\{ \eta_4\;,\,  \xi_2 \}$ can form massive Dirac fermions and decouple.  These are vectorlike  with respect to the  surviving infrared symmetry, (\ref{symbrea}),  thus irrelevant to the anomalies.
  The fermion $\chi_3$  can condense 
\be    \brc  \chi_3 \chi_3 \ckt  \
\ee
forming massive composite mesons, $\sim \chi_3 \chi_3$, which also decouple.  It is again neutral with respect to the unbroken symmetry. 
To summarize,    $SU(4)_c$  is invisible - confines  -  in the IR, and  only the unpaired  part of the $\eta_1$  fermion   $\big({\bar  {\yng(2)}}\big)$  remains  massless.
It's anomalies are reproduced by the composite fermions, (\ref{baryonbra}).

    The massive mesons  $\chi_2\,  \eta_3$,    $\eta_4 \,  \xi_2$,  $\chi_3 \, \chi_3$  are not charged with   the flavor symmetries surviving in the infrared.  
It is tempting to consider them as a sort of  ``dark matter",   as contrasted to  the fermions  $ { B}^{AB} $ which constitute the ordinary, ``visible" sector, in a toy-model interprepretation.

\end{document}